\documentclass[11pt,a4paper]{article}

\RequirePackage[l2tabu, orthodox]{nag}

\usepackage{jheppub}
\usepackage{amsthm,amssymb,amsmath,epic,float}
\usepackage{rotating,epsfig,indentfirst,array,varioref}
\usepackage{appendix,tikz,mathtools}
\usepackage{setspace}
\usepackage{tikz}
\usetikzlibrary{decorations.pathreplacing}
\usetikzlibrary{calc}
\usepackage{enumitem}
\usepackage{hyphenat}
\usetikzlibrary{positioning}
\usepackage{graphicx}  
\usepackage{adjustbox}

\usepackage{arydshln}

\def\be{\begin{equation}}
\def\ee{\end{equation}}

\def\bbt{\bibitem}
\def\be{\begin{equation}}
\def\en{\end{equation}}
\def\ber{\begin{eqnarray}}
\def\enr{\end{eqnarray}}
\def\nmb{ \nonumber\\}
\def\d{\partial}
\def\rbr{\rbrack}
\def\lbr{\lbrack}
\def\rbrc{\rbrace}
\def\lbrc{\lbrace}
\def\ov{\over }
\def\tld{\tilde}

\def\sgm{\sigma}

\def\bet{\beta}

\def\im{\imath}

\def\om{\omega}

\def\eps{\epsilon}

\def\dlt{\delta}
\def\Dl{\Delta}

\def\vl{\vec{l}}
\def\vtl{\tilde{\vec{l}}}
\def\vk{\vec{k}}
\def\vq{\vec{q}}
\def\vt{\vec{t}}
\def\vw{\vec{w}}

\def\vbt{\vec{\beta}}

\usepackage{jheppub}
\makeatletter
\def\@fpheader{\vspace{-.1cm}}
\makeatother

\title{\boldmath  Explicit  construction of $N = 2$ SCFT  orbifold models. \\Spectral flow and mutual locality }

\author[a,c]{Alexander Belavin,}
\author[b]{Vladimir Belavin,}
\author [a]{Sergey Parkhomenko}

\affiliation[a]{Landau Institute for Theoretical Physics, 142432 Chernogolovka, Russia}
\affiliation[b]{Physics Department, Ariel University, Ariel 40700, Israel}
\affiliation[c]{Kharkevich Institute for Information Transmission Problems, 127994 Moscow, Russia}

\emailAdd{belavin@itp.ac.ru}
\emailAdd{vladimirbe@ariel.ac.il}
\emailAdd{spark@itp.ac.ru}

\abstract{In this work we present a new approach to constructing  Calabi-Yau orbifold models  required for compactification in superstring theory. We use the connection of CY orbifolds with the class of exactly solvable N=2 SCFT models to explicitly construct a complete set of fields in these  models using the twisting of the spectral flow and the requirement of mutual locality of the fields.}

\dedicated{In memory of Yaroslav Pugai}


\begin{document} 
\maketitle
\flushbottom

\section{Introduction}
\label{sec:intro}

As is known ten-dimensional superstring theory is a candidate for unification of the Standard Model and quantum gravity.
To get a four-dimensional theory with Space-Time supersymmetry
(which is necessary for phenomenological reasons),  we must compactify the 6 of 10 dimensions on the Calabi-Yau manifold, as shown by Candelas et al. \cite{CHSW}.
Another equivalent approach is the compactification of 6 dimensions into $N=2$ Superconformal field theory with the central charge $c=9$,  as shown by D. Gepner \cite{Gep}.

Each of these two equivalent approaches has its own merits.
Gepner's approach makes it possible to use exactly solvable N=2 SCFT minimal models to construct new ones that are also exactly solvable.
The idea is to use the connection of CY-orbifold models  with the class of exactly solvable minimal $N=2$ SCFT models to explicitly construct a complete set of fields in the models
using Spectral flow transformation \cite {SS} and the requirement of mutual locality of fields.

The large class of CY manifolds  can be defined   
in the weighted projective space
$\mathbb{P}_{n_{1},...,n_{5}}$ in terms of zeros of the quasi-homogeneous polynomials of Fermat type
\ber
W(x_{i}):=\sum_{i=1}^{5}x_{i}^{k_{i}+2}, and \ W(\lambda^{n_i}x_{i})=\lambda^d W(x_{i}),
\ \text{where} \ d= \sum_{i} n_i.
\enr
The last equality, which is the CY condition, can be rewritten as
\ber
\sum_{i=1}^{5}{1\ov k_{i}+2}=1.
\enr
One can also obtain new CY-manifolds as orbifolds of the previous ones by quotienting  them over the so-called admissible group defined as follows
\begin{equation}
\begin{aligned}
G_{adm}\subset G_{tot}=\prod_{i} \mathbb{Z}_{k_{i}+2}, 
\ x_{i}\rightarrow\om_{i}^{w_{i}}x_{i},\ \text{where} \ \om_{i}^{k_{i}+2}=1, \\ 
 \ \sum_{i}{w_{i}\ov k_{i}+2}\in\mathbb{Z}, \
\text{and  the element} \ (1,1,1,1,1)\in G_{adm}
\label{1.Gadm}
\end{aligned}
\end{equation}
It is important that $G_{adm}$ preserves  the product of $x_{i}$, which is equivalent to  preserving a  nonvanishing holomorphic $3$-form on the CY manofold.

The geometric approach to Calabi-Yau of  Fermat class   is equivalent to $N=2$ SCFT approach with $c=9$, realized in terms of the product  $\prod_{i=1}^{5}M_{k_{i}}$,
where $M_{k_{i}}$ is   Minimal model of $N=2$ SCFT with  $c_{i}={3 k_{i}\ov k_{i}+2}$ and
\ber
c= \sum_{i} c_{i}= 9. 
\enr
The last equality coincides with CY condition above.

Thus, Gepner's construction, along with the theory of free fields of transverse degrees of freedom of a superstring in 4-dimensional Minkowski space also contains   a product of $N = 2$ Minimal models. 

In this regard, it makes sense to separate the space-time degrees of freedom from the degrees of freedom of the compact sector and study orbifolds of products of $N = 2$ superconformal minimal models as $N = 2$ $\sgm$-models
of SCFT on Calabi-Yau manifolds. This problem has been discussed in a number of works (see for example \cite{EOTY}), but in our opinion it remains largely unexplored. 
A recent paper \cite{BP} provides some examples of such models built using a new explicit field construction approach to Calabi-Yau  orbifolds of Fermat
type.

In this paper we continue the investigation of the orbifolds of tensor products of $N = 2$ minimal models. 
We present a more general construction of orbifold model fields based on a different point of view than that used in \cite{BP}. Namely, we show how to use such products of exactly solvable
$N=2$ minimal models in order to explicitly construct a set of fields in the orbifold models and thus to obtain new exactly solvable $N=2$ SCFT models with the central charge equal to $9$.

For a few  examples, we construct chiral and anti-chiral rings and show that their dimensions are consistent with the dimensions of the cohomology groups of mutually mirror CY-manifolds, predicted by D. Gepner \cite{DG},
\cite{COGP}   and discovered in \cite{CLS}, \cite{GrPl}, (see also \cite{LS}).

The plan of the paper is as follows. 

In Section~\ref{sec:2} we consider some known  facts about the representation theory 
of the $N=2$ Virasoro superalgebra, $N=(2,2)$ superconfomal field theory models, and $N=(2,2)$ minimal models needed in what follows.
We also present the construction of all primary fields of the minimal models by twisting the spectral flow from chiral primary fields. 
In section~\ref{sec:3} we consider the compositions of the $N=(2,2)$ minimal models and introduce the so called admissible group for the orbifold construction. 
Then we generalize the construction of the spectral flow of primary fields defined in the previous section to the case of composite models
and construct explicitly the complete set of fields of the orbifold model from the requirement of mutual locality.

In section~\ref{sec:4} we prove that OPE of the orbifold model is closed and that all  other conformal bootstrap axioms hold including the modular invariance of the partition function. 
In section~\ref{sec:5} the spectral flow construction (defined in the previous section) is specified to find all $(c,c)$ and $(a,c)$ chiral rings of the model.
In section~\ref{sec:6} we find these rings for several examples of orbifolds and show that they coincide  with corresponding components of the cohomology groups of mutually mirrored CY-manifolds. 
The last fact confirms  the connection  of the constructed orbifold models with the corresponding $\sgm$-models on CY-manifolds.
Section~\ref{sec:7} is a conclusion in which we discuss some open questions.

\section{Preliminaries}
\label{sec:2}

\paragraph{The $N=2$ super Virasoro algebra and its representations.}

Here we recall some general (quite known) properties of $N=2$ Virasoro superalgebra and its representations, which are necessary for our further construction.

 The commutation relations for the generators $L_n, J_{n}, G^{\pm}_{r}$ 
of $N=2$ Super-Virasoro algebra have the form

\begin{equation}
\begin{aligned}
&\lbr  L_{n},L_{m}\rbr=(n-m)L_{n+m}+{c\ov 12}(n^{3}-n)\dlt_{n+m,0},\\
&\lbr J_{n},J_{m}\rbr={c\ov 3}n\dlt_{n+m,0},\\
&\lbr L_{n},J_{m}\rbr=-mJ_{n+m},\\
&\lbrc G^{+}_{r},G^{-}_{s}\rbrc=L_{r+s}+{r-s\ov 2}J_{r+s}+{c\ov 6}(r^{2}-{1\ov 4})\dlt_{r+s,0},\\
&\lbr L_{n},G^{\pm}_{r}\rbr=({n\ov 2}-r)G^{\pm}_{r+n},\\
&\lbr J_{n},G^{\pm}_{r}\rbr=\pm G^{\pm}_{r+n}.
\label{2.N2Vir}
\end{aligned}
\end{equation}

The representations of the $N = 2$ superalgebra split into two sectors. In the first sector, which is called the Nevew-Schwartz sector ($NS$)
the values of the modes $r, s$ of the fermionic generators are half-integers. In the second so called Ramond sector ($R$) the modes $r, s$ of the fermionic currents take integer values. 

Note that the second and third commutation relations mean that $J_{n}$ are the Fourier modes of the free bosonic $U(1)$-current $J(z)$, 
which can be written as
\ber
J(z)=\im\sqrt{c\ov 3}{\d\phi\ov \d z},
\label{2.Jboson}
\enr
where $\phi(z)$ is free massles scalar field.

In both sectors there exist highest weight representations.
In the $NS$ sector, the highest weight representation is generated from the highest weight vector $\Phi^{NS}_{\Dl,Q}$, which is called the primary state in standard terminology.
This state is determined by the relations
\begin{equation}
\begin{aligned}
&L_{n}\Phi^{NS}_{\Delta,Q}=0, \ J_{n}\Phi^{NS}_{\Delta,Q}=0, \ n\geq 1,
\
G^{\pm}_{r}\Phi^{NS}_{\Delta,Q}=0, \ r\geq {1\ov 2},
\\
&\qquad\qquad L_{0}\Phi_{(\Delta,Q)}=\Delta \Phi_{(\Delta,Q)},
\
J_{0}\Phi_{(\Delta,Q)}=Q\Phi_{(\Delta,Q)}.
\label{2.NSPrim}
\end{aligned}
\end{equation}

Thus, $\Dl$ and $Q$ are the conformal dimension and $U(1)$ charge of the primary state $\Phi^{NS}_{\Dl,Q}$. 
The descendant states are generated from the primary state above by the creation generators of $N=2$ Virasoro superalgebra.
 We form in this way  a Verma module which is reducible in general. 
The corresponding irreducible highest weight representation ${\cal{H}}^{NS}_{\Dl,Q}$ is given by a quotient of the Verma module by its submodules.

The primary state (highest weight vector) $\Phi^{R}_{\Delta,Q}$ in the $R$ sector has similar properties, but the annihilation conditions for fermionic generators are slightly different:
\ber
G^{\pm}_{n}\Phi^{R}_{\Delta,Q}=0, \ n \geq 1,\quad
G^{+}_{0}\Phi^{R}_{\Delta,Q}=0.
\label{2.RPrim}
\enr
Applying to the state $\Phi^{R}_{\Dl,Q}$ the creation generators of $N=2$ Virasoro superalgebra we form a Verma module. The corresponding highest weight representation ${\cal{H}}^{R}_{\Dl,Q}$ is certain quotient of the Verma module by its submodules.
 The structure of Verma modules of $N=2$ Virasoro superalgebra has been analyzed in \cite{VPY}.

A special class of representations in $NS$ sector arise for primary states which are subject to an additional constraint
\ber
G^{+}_{-{1\ov 2}}\Phi^{NS}_{\Delta,Q}=0.
\label{2.CPrim}
\enr
Such states are called chiral primary states. 

 Similarly, if the primary state, in addition to (\ref{2.NSPrim}), satisfies
\ber
G^{-}_{-{1\ov 2}}\Phi^{NS}_{\Delta,Q}=0,
\label{2.APrim}
\enr
then it is called anti-chiral primary state.

 Chiral and anti-chiral primary states, which were first discussed in \cite{LVW}, have special properties. Namely, due to the anti-commutators 
from (\ref{2.N2Vir}) the dimension and charge of the chiral primary state are not independent:
\ber
Q=2\Delta.
\label{2.CPrim1}
\enr
For the anti-chiral primary state
\ber
Q=-2\Delta.
\label{2.APrim1}
\enr

It is possible to continuously connect $NS$ sector to the $R$ sector using one-parametric family of automorphisms of $N=2$ Virasoro superalgebra 
\cite{LVW} known as spectral flow:
\begin{equation}
\begin{aligned}
&\tilde{G}^{\pm}_{r}=U^{-t}G^{\pm}_{r}U^{t}=G^{\pm}_{r\pm t},\\
&\tilde{J}_{n}=U^{-t}J_{n}U^{t}=J_{n}+{c\ov3}t\dlt_{n,0},\\
&\tilde{L}_{n}=U^{-t} L_{n} U^{t}=L_{n}+tJ_{n}+{c\ov 6}t^{2}\dlt_{n,0}.
\label{2.Sflow}
\end{aligned}
\end{equation}
So, one can see that for $t\in \mathbb{Z}+{1\ov 2}$ the spectral flow interpolates between $NS$ and $R$ sectors and for $t\in\mathbb{Z}$ it takes $NS$ to $NS$ and $R$ to $R$. It makes sense to emphasize that as a result of the  spectral flow transformation , we obtain a new, so-called twisted representation
\ber
{\cal{H}}^{t}_{\Delta,Q}=U^{t}{\cal{H}}_{\Delta,Q}.
\enr
It allows in particular to define $R$ sector representations as the ${1\ov 2}$-twisted $NS$ sector representations:
\ber
{\cal{H}}^{R}_{\Delta_{R},Q_{R}}=U^{{1\ov 2}}{\cal{H}}^{NS}_{\Delta_{NS},Q_{NS}},
\label{2.HR}
\enr
where
\ber
\Delta_{R}=\Delta_{NS}+{Q_{NS}\ov 2}+{c\ov 24}, Q_{R}=Q_{NS}+{c\ov 6}.
\label{2.HRDQ}
\enr
For the construction that will be discussed in what follows it is important that not only $U(1)$ current can be expressed by the free scalar (\ref{2.Jboson}), but the generator of the spectral flow $U$ is also expressed in terms of the free scalar field $\phi(z)$:
\ber
U=\exp{(\im \sqrt{c\ov 3}\phi(z))}.
\label{2.Uboson}
\enr

\paragraph{Minimal representations.}
In the case when  the central charge takes one of the values
\ber
c={3k\ov k+2}, \ k=0,1,2,....
\label{2.cMin}
\enr
there exists a finite set of irreducible integrable unitary representations of $N=2$ super Virasoro, numerated by pairs of numbers $l, q$
\ber
{\cal{H}}^{NS}_{l,q}, \ {\cal{H}}^{R}_{l,q}, \
l=0,1,...,k, \ q=-l,-l+2,...,l.
\label{2.MinRep}
\enr 
We will call them minimal representations.
The conformal dimensions and charges of the corresponding primary 
states $\Phi^{NS}_{l,q}$ in $NS$ sector are
\ber
\Delta_{l,q}=\frac{l(l+2)-q^{2}}{4(k+2)}, \ \ Q_{l,q}=\frac{q}{k+2}.
\label{2.DeltQ}
\enr
In $R$ sector the dimensions and charges of primary states $\Phi^{R}_{l,q}$ are given by
\ber
\Delta^{R}_{l,q}=\frac{l(l+2)-(q-1)^{2}}{4(k+2)}+\frac{1}{8},\quad
\bar{Q}^{R}_{\l,q}=Q^{R}_{l,q}=\frac{q-1}{k+2}+\frac{1}{2}.
\label{2.DeltQR}
\enr
Among the primary states the  chiral primary states appear when $q=l$:
\ber
\Phi^{c}_{l}\equiv \Phi^{NS}_{l,l}, \quad \Phi^{a}_{l}\equiv \Phi^{NS}_{l,-l},
\label{2.CPrim}
\enr
while anti-chiral primary states appear when $q=-l$:
\ber
\Phi^{a}_{l}\equiv \Phi^{NS}_{l,-l}.
\label{2.APrim}
\enr

\paragraph{Spectral flow construction of primary states for minimal representations.}

As we already mentioned the spectral flow automorphism allows one to relate $NS$ representations to  $R$ representations. 
For the minimal representations (\ref{2.MinRep}) the spectral flow play much more prominent role \cite{FST}. 
We now show that all primaries from (\ref{2.MinRep}) can be explicitly constructed from chiral primaries (\ref{2.CPrim}) using spectral flow automorphism (\ref{2.Sflow}).

We start by the spectral flow construction an anti-chiral primary state. It uses   the so called extremal vector of $N=2$ Virasoro superalgebra representation. 
The extremal vector we are dealing with is a special descendant state in the irreducible representation of $N=2$ Virasoro superalgebra generated from the chiral primary state $\Phi^{c}_{l}$ as
\begin{equation}
\begin{aligned}
&E^{-}_{l}=G^{-}_{{1\ov 2}-l}...G^{-}_{-{1\ov 2}}\Phi^{c}_{l},
\\
&G^{+}_{r}E^{-}_{l}=0, \,\, \ r\geq {1\ov 2}+l, \quad
G^{-}_{r}E^{-}_l=0, \ \ r\geq-{1\ov 2}-l, 
\\
&J_{n}E^{-}_l=L_{n}E^{-}_l=0, \ \ n\geq 1.
\label{3.Ext}
\end{aligned}
\end{equation}

The extremal vector $E^{-}_l$ is associated with one of the possible Borel subalgebras in the $N=2$ Super-Virasoro algebra  generated by
\ber
G^{+}_{r}, r\geq {1\ov 2}+l, \quad
G^{-}_{r}, r\geq -{1\ov 2}-l, \quad
 J_{n}, \ L_{n}, n\geq 0.
\enr
The vector $E^{-}_l$  is isomorphic to the anti-chiral  primary 
state $\Phi^{a}_{l}$. 
The isomorphism is given by the following spectral flow construction
\ber
U^{l}E^{-}_{l}=\Phi^{a}_{l}.
\label{3.AC}
\enr
This two-stage procedure can be applied to construct all primary states of the $NS$ sector.
Let us consider an example  of the descendant state $G^{-}_{-{1\ov 2}}\Phi^{c}_{l}$.
By direct calculations one can show that the state
\ber
UG^{-}_{-{1\ov 2}}\Phi^{c}_{l}
\label{3.Priml1}
\enr
gives the spectral flow realization of the primary state $\Phi_{l,l-2}$. 

More generally, the state
\ber
\Phi_{l,t}=U^{t}G^{-}_{-t+{1\ov 2}}...G^{-}_{-{1\ov 2}}\Phi^{c}_{l}, \quad 0\leq t\leq l.
\label{2.Primlt}
\enr
gives spectral flow realization of the primary state $\Phi_{l,q}$, where $q=l-2t$. Using (\ref{2.Sflow}) this expression can be represented in the following form
\ber
\Phi_{l,t}=(UG^{-}_{-{1\ov 2}})^{t}\Phi^{c}_{l}, \quad 0\leq t\leq l.
\label{2.Prim1}
\enr
For the purposes of the orbifold construction we need to extend this formula. 
Namely, we define the vector
\ber
\Phi_{l,t}=(UG^{-}_{-{1\ov 2}})^{t-l-1}U(UG^{-}_{-{1\ov 2}})^{l}\Phi^{c}_{l}, \quad l+1\leq t\leq k+1.
\label{2.Prim2}
\enr
It can be checked that  the state $\Phi_{l,t}$ gives the spectral flow realization of the primary state $\Phi_{\tld{l},\tld{q}}$, where $\tld{l}=k-l$, $\tld{q}=k+2+l-2t$. 

In fact, this formula can be extended. Indeed, if consider  the vector
\ber
U(UG^{-}_{-{1\ov 2}})^{k-l}U(UG^{-}_{-{1\ov 2}})^{l}\Phi^{c}_{l}(z),
\enr
then it is easy to verify that in this way we obtain another realization of the initial chiral primary state $\Phi^{c}_{l}$.
 Thus, making $k+2$ steps, we return to the original representation of $N=2$ Virasoro superalgebra.
 So, for the spectral flow the following periodicity property holds
\ber
U^{k+2}\approx 1.
\label{2.Period}
\enr 
In what follows we will use the spectral flow parameter $t$ instead of $q$ in order to stress the spectral flow realization of primary states
(\ref{2.Prim1}), (\ref{2.Prim2}).

The construction of the spectral flow of the primary  states in the $R$-sector can  be obtained by applying the operator $U^{1\ov 2}$ to the expressions (\ref{2.Prim1}), (\ref{2.Prim2}).
\vskip .3 cm

\paragraph{The characters.}
As was shown in \cite{Gep} it is convenient to split the space of each irreducible representation of $N=2$ algebra into two subspace that are representations  of the subalgebra generated by an even number of the fermionic generators $G^{\pm}$.

Then for the minimal representations (\ref{2.MinRep}) we can define  the characters
\begin{equation}
\begin{aligned}
&NS_{l,q}(\tau,\theta,\eps)=\exp{(-\im 2\pi\eps)}Tr_{{\cal{H}}^{NS}_{l,q}}
(\exp{(\im 2\pi (L_{0}-{c\ov 24})\tau+\im 2\pi J_{0}\theta)}),\\
&\tld{NS}_{l,q}(\tau,\theta,\eps)=\exp{(-\im 2\pi\eps)}Tr_{{\cal{H}}^{NS}_{l,q}}((-1)^{F}\exp{(\im 2\pi (L_{0}-{c\ov 24})\tau+\im 2\pi J_{0}\theta)}),\\
&R_{l,q}(\tau,\theta,\eps)=\exp{(-\im 2\pi\eps)}Tr_{{\cal{H}}^{R}_{l,q}}(\exp{(\im 2\pi (L_{0}-{c\ov 24})\tau+\im 2\pi J_{0}\theta)}),\\
&\tld{R}_{l,q}(\tau,\theta,\eps)=\exp{(-\im 2\pi\eps)}Tr_{{\cal{H}}^{R}_{l,q}}((-1)^{F}\exp{(\im 2\pi (L_{0}-{c\ov 24})\tau+
\im 2\pi J_{0}\theta)}),
\label{2.charac}
\end{aligned}
\end{equation}
where $(-1)^{F}$ is fermionic number operator.

The modular transformation properties of characters, which we  use below,  were obtained in \cite{Gep}, look as follows:
\begin{equation}
\begin{aligned}
&NS_{l,q}(\tau+1,\theta,\eps)=
\exp{(\im 2\pi(\Delta_{l,q}-{c\ov 24}))}\tld{NS}_{l,q}(\tau,\theta,\eps),
\\
&\tld{NS}_{l,q}(\tau+1,\theta,\eps)=\exp{(\im 2\pi(\Delta_{l,q}-{c\ov 24}))}NS_{l,q}(\tau,\theta,\eps),
\\
&R_{l,q}(\tau+1,\theta,\eps)=
\exp{(\im 2\pi(\Delta^{R}_{l,q}-{c\ov 24}))}R_{l,q}(\tau,\theta,\eps),
\\
&\tld{R}_{l,q}(\tau+1,\theta,\eps)=
\exp{(\im 2\pi(\Delta^{R}_{l,q}-{c\ov 24}))}\tld{R}_{l,q}(\tau,\theta,\eps),
\\
&NS_{l,q}(-{1\ov\tau},{\theta\ov\tau},\eps+{c\theta^{2}\ov 6\tau})=
\sum_{l',q'}S^{l',q'}_{l,q}NS_{l',q'}(\tau,\theta,\eps),
\\
&\tld{NS}_{l,q}(-{1\ov\tau},{\theta\ov\tau},\eps+{c\theta^{2}\ov 6\tau})=
\sum_{l',q'} S^{l',q'-1}_{l,q}R_{l',q'}(\tau,\theta,\eps), 
\\
&R_{l,q}(-{1\ov\tau},{\theta\ov\tau},\eps+{c\theta^{2}\ov 6\tau})=
\sum_{l',q'}S^{l',q'}_{l,q-1}\tld{NS}_{l'q'}(\tau,\theta,\eps),
\\
&\tld{R}_{l,q}(-{1\ov\tau},{\theta\ov\tau},\eps+{c\theta^{2}\ov 6\tau})=
-\im\sum_{l',q'}S^{l',q'-1}_{l,q-1}\tld{R}_{l',q'}(\tau,\theta,\eps),
\label{2.STact}
\end{aligned}
\end{equation}
where the S-matrix 
\ber
S^{l',q'}_{l,t}=\sin{(\pi{(l+1)(l'+1)\ov k+2})}\exp{(\im\pi{qq'\ov k+2})}.
\label{2.Smatr}
\enr
It follows that the characters of the $N=2$ minimal model form a unitary representation of the modular group \cite{Gep}.
\vskip .3 cm

\paragraph{Models of $N=(2,2)$  Superconformal Field Theory.}

Like any models of CFT, the $N =(2,2)$ superconformal models are characterized by two copies of their symmetry algebra.

The space of states of $N =(2,2)$ superconformal field theory consists of the local fields, which form the products of representations of holomorphic and anti-holomorphic $N=2$ Virasoro superalgebras. So the total space of states can be written in the form
\begin{equation}
\begin{aligned}
{\cal{H}}={\cal{H}}^{NS}\otimes\bar{\cal{H}}^{NS}\oplus {\cal{H}}^{R}\otimes\bar{\cal{H}}^{R},
\label{2.Htot}
\end{aligned}
\end{equation}
where
\begin{equation}
\begin{aligned}
&{\cal{H}}^{NS}\otimes\bar{\cal{H}}^{NS}=\oplus_{\Delta,\bar{\Delta}}\oplus_{Q,\bar{Q}}{\cal{H}}^{NS}_{\Delta,Q}\otimes\bar{\cal{H}}^{NS}_{\bar{\Delta},\bar{Q}},
\\
&{\cal{H}}^{R}\otimes\bar{\cal{H}}^{R}=\oplus_{\Delta,\bar{\Delta}}\oplus_{Q,\bar{Q}}{\cal{H}}^{R}_{\Delta,Q}\otimes\bar{\cal{H}}^{R}_{\bar{\Delta},\bar{Q}}
\label{2.HNSR}
\end{aligned}
\end{equation}
and ${\cal{H}}^{NS}_{\Delta,Q}$ ($\bar{\cal{H}}^{NS}_{\bar{\Delta},\bar{Q}}$) is a highest weight $(\Delta,Q)$ ($\bar{\Delta},\bar{Q}$) representation of the holomorphic (anti-holomorphic) $N=2$ superalalgebra Virasoro in $NS$($\overline{NS}$) sector considered above. The similar meaning has the expansion for ${\cal{H}}^{R}\otimes\bar{\cal{H}}^{R}$. 

The primary fields $\Psi_{(\Delta,Q)(\bar{\Delta},\bar{Q})}(z,\bar{z})$ are the primary states w.r.t the holomorphic and anti-holomorphic $N=2$ Virasoro superalgebras. These fields can be considered as a proper products of holomorphic and anti-holomorphic factors:
\ber
\Psi_{(\Delta,Q)(\bar{\Delta},\bar{Q})}(z,\bar{z})=\Phi_{\Delta,Q}(z)\bar{\Phi}_{\bar{\Delta},\bar{Q}}(\bar{z}).
\label{2.Primf}
\enr
The rest local fields are generated from the primary fields by applying creation generators of holomorphic and anti-holomorphic $N=2$ Virasoro superalgebras. OPE of the local fields have to be  closed and associative as required by Conformal Bootstrap axioms \cite{BPZ}.

Among the primary fields there are special ones, which corresponds to the chiral and anti-chiral primary states. Thus, in general $N=(2,2)$ superconformal model there are 4 types of special primary fields depending on whether their holomorphic and anti-holomorphic factors are the chiral or anti-chiral states:
\begin{equation}
\begin{aligned}
&\Psi^{cc}_{\Delta,\bar{\Delta}}(z,\bar{z})=\Phi^{c}_{\Delta}(z)\bar{\Phi}^{c}_{\bar{\Delta}}(\bar{z}),
\
\Psi^{ac}_{\Delta,\bar{\Delta}}(z,\bar{z})=\Phi^{a}_{\Delta}(z)\bar{\Phi}^{c}_{\bar{\Delta}}(\bar{z}),\
\\
&\Psi^{ca}_{\Delta,\bar{\Delta}}(z,\bar{z})=\Phi^{c}_{\Delta}(z)\bar{\Phi}^{a}_{\bar{\Delta}}(\bar{z}),
\
\Psi^{aa}_{\Delta,\bar{\Delta}}(z,\bar{z})=\Phi^{a}_{\Delta}(z)\bar{\Phi}^{a}_{\bar{\Delta}}(\bar{z}).
\label{2.Rings}
\end{aligned}
\end{equation}

These fields form 4 rings in the general $N=(2,2)$ SCFT, as shown in \cite{LVW}.
In the geometric approach, in which $N=2$ SCFT is interpreted as a $\sigma$ -models on CY manifolds, these rings are related to the cohomology classes of CY-manifold.


\vskip 10pt
A minimal model $M_{k}$ of  SCFT with $N=(2,2)$ symmetry,  is labeled by positive integer $k$. Its space of local fields can be decomposed into the sum of products of minimal representations (\ref{2.MinRep}). 
In general, the set of local fields of the minimal models have an $A-D-E$ classification, but in this paper we consider only $A$ series.  Then the primary fields of the model are given by diagonal pairing of holomorphic and anti-holomorphic factors:
\begin{equation}
\begin{aligned}
&\Psi^{NS}_{l,q}(z,\bar{z})=\Phi^{NS}_{l,q}(z)\bar{\Phi}^{NS}_{l,q}(\bar{z}),
\\
&\Psi^{R}_{l,q}(z,\bar{z})=\Phi^{R}_{l,q}(z)\bar{\Phi}^{R}_{l,q}(\bar{z}),
\\
& \text{where}\quad l=0,...,k, \quad q=-l,-l+2,...,l.
\label{2.PrimNS}
\end{aligned}
\end{equation}
The factors $\Phi^{NS}_{l,q}(z)$ $\bar{\Phi}^{NS}_{l,q}(\bar{z})$, $\Phi^{R}_{l,q}(z)$, $\bar{\Phi}^{R}_{l,q}(\bar{z})$ are the primary states of the minimal representations (\ref{2.MinRep})
so that their conformal dimensions and $U(1)$ charges  are given by (\ref{2.DeltQ}) and
(\ref{2.DeltQR}) respectively.

Because of (\ref{2.CPrim}), (\ref{2.APrim}) we see that in the Minimal model there are only chiral-chiral primary fields
\ber
\Psi^{cc}_{l}(z,\bar{z})=\Phi^{c}_{l}(z)\bar{\Phi}^{c}_{l}(\bar{z}),
\nmb
\enr
and antichiral-antichiral primary fields
\ber
\Psi^{aa}_{l}(z,\bar{z})=\Phi^{a}_{l}(z)\bar{\Phi}^{a}_{l}(\bar{z})
\nmb
\enr
All diagonally composed fields  from $NS$ sector are mutually local.

The $NS$ sector fields are mutually quasi-local with $R$ sector fields, i.e.,  when one field goes around the other, there is the phase factor $(-1)^{F+\bar{F}}$, where $F$ and $\bar{F}$ are the fermionic numbers of the $NS$ field.
The $R$ sector fields are mutually quasi-local, i.e., when one field goes around the other, there is the phase factor $(-1)^{F_{1}+\bar{F}_{1}+F_{2}+\bar{F}_{2}}$, where $F_{1,2}$ and $\bar{F}_{1,2}$ are the fermionic numbers of the $R$ fields.
This mutual quasi-locality structure is a particular case of the general picture valid for any $N=(1,1)$ superconformal theory.

From the total set of fields of the minimal model $M_{k}$ one can single out the subset of mutually local fields~\cite{Gep}:
\ber
{\cal{H}}_{loc}={\cal{H}}^{NS}_{0}\oplus {\cal{H}}^{NS}_{2}\oplus{\cal{H}}^{R}_{1}
\oplus{\cal{H}}^{R}_{3},
\label{2.Local}
\enr
where ${\cal{H}}^{NS}_{0}$ is the subspace of ${\cal{H}}^{NS}\otimes\bar{\cal{H}}^{NS}$with even fermion numbers $F$ and $\bar{F}$, ${\cal{H}}^{NS}_{2}$ is the subspace of ${\cal{H}}^{NS}\otimes\bar{\cal{H}}^{NS}$ with odd fermion numbers $F$ and $\bar{F}$, the spaces ${\cal{H}}^{R}_{1,3}$ have similar meaning for $R$ sector fields. 

Then, the  partition function calculated over this subspace
\be
\begin{aligned}
Z(\tau,\bar{\tau},0,0)=\sum_{l,q}(NS_{l,q}(\tau,0)NS^{*}_{l,q}(\bar{\tau},0)+
\tld{NS}_{l,q}(\tau,0)\tld{NS}^{*}_{l,q}(\bar{\tau},0)+
\\
R_{l,q}(\tau,0)R^{*}_{l,q}(\bar{\tau},0)+
\tld{R}_{l,q}(\tau,0)\tld{R}^{*}_{l,q}(\bar{\tau},0))
\label{2.ZMin}
\end{aligned}
\ee
is modular invariant due to (\ref{2.STact}) (see \cite{Gep}).

Thus, we obtain a QFT model that satisfies all the axioms of a conformal bootstrap.

\paragraph{The composition of $N=(2,2)$ Minimal Models.}

We now construct, following \cite{Gep}, a set of fields of the $N=(2,2)$ SCFT model with the total central charge $9$, starting from the composition of five minimal models.

The conformal symmetry of the model is the Virasoro superalgebra $N=(2,2)$, defined as the diagonal subalgebra in the tensor product of 5 minimal models as follows
\ber
L_{tot,n}=\sum_{i}L_{(i),n}, \ J_{tot,n}=\sum_{i}J_{(i),n},
\
G^{\pm}_{tot,r}=\sum_{i}G^{\pm}_{(i),r}.
\label{2.DiagVir}
\enr

The action of this algebra is correctly defined only on the product of $NS$ representations  or on the product of  $R$  representations of minimal models $M_{k_{i}}$.
Therefore, we can form local $NS$ or $R$ primary fields  in the product model $M_{\vk}$ by taking only products of primary fields from each minimal model $M_{k_{i}}$  belonging to  the same ($ NS$ or $R$) sectors:

\be
\begin{aligned}
&\Psi^{NS}_{\vec{l},\vec{q}}(z,\bar{z})=\prod_{i}\Psi^{NS}_{l_{i},q_{i}}(z,\bar{z}),
\\
&\Psi^{R}_{\vl,\vq}(z,\bar{z})=\prod_{i}\Psi^{R}_{l_{i},q_{i}}(z,\bar{z})=\prod_{i}U_{i}^{1\ov 2}\bar{U}_{i}^{1\ov 2}\Psi^{NS}_{l_{i},q_{i}}(z,\bar{z}).
\label{2.Prim}
\end{aligned}
\ee

The descendant fields are generated from these primary fields  by the creation generators of the $N=2$ Virasoro superalgebras of the $M_{k_{i}}$ models. It is clear that the OPE of the fields constructed by this way is associative and closed. It follows then from the minimal model considerations that $NS$ sector fields of the product model are mutually local, while $R$ sector fields are mutually quasi-local among themselves and with $NS$ sector fields.

One can construct the subspace of mutually local fields in the product model similarly to the minimal model case \cite{Gep}. The fields are mutually local if their total fermion numbers $F$ and $\bar{F}$ are equal. It gives the decomposition of the total space of mutually local fields into the 4 subspaces:
\ber
{\cal{H}}_{loc}={\cal{H}}^{NS}_{0}\oplus {\cal{H}}^{NS}_{2}\oplus{\cal{H}}^{R}_{1}
\oplus{\cal{H}}^{R}_{3},
\label{2.ProdLocal}
\enr
where ${\cal{H}}^{NS}_{0}$ is the subspace of ${\cal{H}}^{NS}\otimes\bar{\cal{H}}^{NS}$with even total fermion numbers $F$ and $\bar{F}$, ${\cal{H}}^{NS}_{2}$ is the subspace of ${\cal{H}}^{NS}\otimes\bar{\cal{H}}^{NS}$ with odd total fermion numbers $F$ and $\bar{F}$, the spaces ${\cal{H}}^{R}_{1,3}$ have similar meaning for $R$ sector fields.

Similar to the minimal model case one can build the modular invariant partition function getting contributions only from the subspace ${\cal{H}}_{loc}$. To this end we have to consider the aligned products of characters:
\be
\begin{aligned}
&NS_{\vl,\vq}(\tau,\theta,\eps)=\prod_{i}NS_{l_{i},q_{i}}(\tau,\theta,\eps),
\
\widetilde{NS}_{\vl,\vq}(\tau,\theta,\eps)=\prod_{i}\widetilde{NS}_{l_{i},q_{i}}(\tau,\theta,\eps),
\\
&R_{\vl,\vq}(\tau,\theta,\eps)=\prod_{i}R_{l_{i},q_{i}}(\tau,\theta,\eps),
\
\widetilde{R}_{\vl,\vq}(\tau,\theta,\eps)=\prod_{i}\widetilde{R}_{l_{i},q_{i}}(\tau,\theta,\eps).
\label{2.charac}
\end{aligned}
\ee
Then one check directly the modular invariance of the partition function \cite{Gep}
\be
\begin{aligned}
Z(\tau,\bar{\tau})_{prod}=\sum_{\vl,\vq}(NS_{\vl,\vq}(\tau,0)NS^{*}_{\vl,\vq}(\bar{\tau},0)+
\tld{NS}_{\vl,\vq}(\tau,0)\tld{NS}^{*}_{\vl,\vq}(\bar{\tau},0)+
\\
R_{\vl,\vq}(\tau,0)R^{*}_{\vl,\vq}(\bar{\tau},0)+
\tld{R}_{\vl,\vq}(\tau,0)\tld{R}^{*}_{\vl,\vq}(\bar{\tau},0)).
\label{2.ZProdMin}
\end{aligned}
\ee


\section{Construction of $N=(2,2)$ orbifods} 
\label{sec:3}

 In this section, we give the construction of new $N=(2,2)$ SCFT models with the total central charge $9$, starting from the compositions of the $M_k$ models discussed above and using the so-called admissible group.

To define the admissible group \cite{BH} we recall that 
the composition of the $M_k$ models  model has a discrete symmetry group which is defined as follows
\ber
G_{tot}=\prod_{i=1}^{5}\mathbb{Z}_{k_{i}+2}=\lbrc \prod_{i} \hat{g}_{i}^{w_{i}}, \ w_{i}\in \mathbb{Z}, 
\hat{g}_{i}=\exp{(\im 2\pi J_{(i),0})}\rbrc.
\label{3.Gtot}
\enr

The admissible group is a subgroup of $G_{tot}$ which is defined as follows:
\be
\begin{aligned}
G_{adm}=\lbrc &\vw=\sum_{a=0}^{K-1}m_{a}\vbt_{a}, \ m_{a}\in\mathbb{Z},
\
\sum_{i}{\bet_{ai}\ov k_{i}+2}\in\mathbb{Z}, \ \bet_{ai}\in\mathbb{Z},
\\
&\vbt_{0}=(1,1,1,...,1),\ \sum_{i}{1\ov k_{i}+2}=1\rbrc\subset G_{tot}. 
\label{3.Gadm}
\end{aligned}
\ee
Thus, the 5-dimensional vectors $\vbt_{a}$ are the generators of the admissible group, so that an arbitrary element $\vw\in G_{adm}$ can be decomposed in terms of these generators.
Notice that $G_{adm}$ coincides with the group that preserves the nowhere vanishing holomorphic $(3,0)$-form  
 on CY manifold defined as a hypesurfuce  in $\mathbb{P}_{n_{1},...,n_{5}}$.

We use also the spectral flow construction of primary fields, 
similar to the one used above (\ref{2.Prim1}),(\ref{2.Prim2}) in the case of a single minimal model,  as well as  the requirement of mutual locality for constructing all the fields in the orbifold model.

In the first step, we use the elements $\vw$ of the admissible group (\ref{3.Gadm})
for expanding the state space of the product model $M_{\vk}$ by adding in $NS$ sector twisted fields of the form
\ber
\Psi^{NS}_{\vl,\vt,\vw}(z,\bar{z})=V_{\vl,\vt+\vw}(z)\bar{V}_{\vl,\vt}(\bar{z}),
\label{3.Primo}
\enr
where
\ber
\bar{V}_{\vl,\vt}(\bar{z})=\prod_{i}\bar{V}_{l_{i},t_{i}}(\bar{z}),\
\bar{V}_{l_{i},t_{i}}(z)=(UG^{-}_{-{1\ov 2}})_{i}^{t_{i}}\bar{\Phi}^{c}_{l_{i}}(z), \ 0\leq t_{i}\leq l_{i},
\label{3.Primo1}
\enr
and
\be
V_{\vl,\vt+\vw}(z)=\prod_{i=1}^{5}V_{l_{i},t_{i}+w_{i}}(z),
\ee
where
\be
\begin{aligned}
&V_{l_{i},t_{i}+w_{i}}(z)=\begin{cases}(UG^{-}_{-{1\ov 2}})_{i}^{t_{i}+w_{i}}\Phi^{c}_{l_{i}}(z), 
\qquad \text{if}\quad \ 0\leq t_{i}+w_{i}\leq l_{i}, \\\\
(UG^{-}_{-{1\ov 2}})_{i}^{t_{i}+w_{i}-l_{i}-1}
U_{i}(UG^{-}_{-{1\ov 2}})_{i}^{l_{i}}\Phi^{c}_{l_{i}}(z), 
\quad \text{if}\quad \ l_{i}+1\leq t_{i}+w_{i}\leq k_{i}+1.
 \end{cases}
\label{3.Primo2}
\end{aligned}
\ee

In the second step, we require the mutual locality of the fields obtained above. To  derive the condition which follows from this,
we consider two such fields
\be
\begin{aligned}
&\Psi^{NS}_{\vl_{1},\vt_{1},\vw_{1}}(z,\bar{z}), \quad \Psi^{NS}_{\vl_{2},\vt_{2},\vw_{2}}(0,0),
\\
&\vw_{1}=w^{a}_{1}\vbt_{a}, \ \vw_{2}=w^{a}_{2}\vbt_{a}\in G_{adm}.
\end{aligned}
\ee
and calculate the phase $\exp{(\im\theta)}$ that occurs when the first field goes around the second.

In order to make this calculation explicitly we use the bosonic representation for the spectral flow operators 
$U_{i}$ and for the currents $J_{(i)}(z)$ for each minimal model 
$M_{k_{i}}$. We do this by introducing free bosonic fields $\phi_{i }(z)$, $ i = 1,...,5$, whose  OPE has the form
\ber
\phi_{i}(z)\phi_{j}(0)=-\dlt_{ij}\log{(z)}+\cdots\, .
\label{3.UJbos}
\enr
Then
\be
\begin{aligned}
&J_{i}(z)=\im\sqrt{k_{i}\ov k_{i}+2}\d\phi_{i}(z), \quad U_{i}(z)=\exp{(\im\sqrt{k_{i}\ov k_{i}+2}\phi_{i}(z))},
\\
&\qquad\Phi_{l_{i},t_{i}}(z)=\exp{(\im{q_{i}\ov\sqrt{k_{i}(k_{i}+2)}}\phi_{i}(z))}\hat{\Phi}_{l_{i},t_{i}}(z),
\label{3.UPhi}
\end{aligned}
\ee
where $q_{i}=l_{i}-2t_{i}$ and $\hat{\Phi}_{l_{i},t_{i}}(z)$ is neutral w.r.t. the current $J_{i}(z)$.

Since all the fields in $NS$ sector of the product model $M_{\vk}$ are mutually local, we only  need to calculate the phase factor in the operator product
\ber
\exp{[\im\sum_{i}(\sqrt{k_{i}\ov k_{i}+2}w_{1i}+{q_{1i}\ov\sqrt{k_{i}(k_{i}+2)}})\phi_{i}(z)]}
\exp{[\im\sum_{j}(\sqrt{k_{j}\ov k_{j}+2}w_{2j}+{q_{2j}\ov\sqrt{k_{j}(k_{j}+2)}})\phi_{j}(0)]}.
\nmb
\label{3.VOpe}
\enr
The result is
\be
\begin{aligned}
&\exp{(\im\theta)}=
\\
&\quad\exp{(\im 2\pi\sum_{a=0}^{K-1}w^{a}_{1}(Q^{\vbt_{a}}_{\vl_{2},\vt_{2}}-
\sum_{b=0}^{K-1}w^{b}_{2}\sum_{i}{\bet_{ai}\bet_{bi}\ov k_{i}+2}))}
\exp{(\im 2\pi\sum_{a=0}^{K-1}w^{a}_{2}(Q^{\vbt_{a}}_{\vl_{1},\vt_{1}}-
\sum_{b=0}^{K-1}w^{b}_{1}\sum_{i}{\bet_{ai}\bet_{bi}\ov k_{i}+2}))},
\label{3.Phase}
\end{aligned}
\ee
where
\ber
Q^{\vbt_{a}}_{\vl,\vt}=\sum_{i}{\bet_{ai}q_{i}\ov k_{i}+2}.
\label{3.Qbet}
\enr

Mutual locality means that for each arbitrary pair of twisted sectors $\vw_{1}$ and $\vw_{2}$ the phase factor above must be trivial. 
Therefore, the field $\Psi^{NS}_{\vl,\vt,\vw}(z,\bar{z})$ will be mutually local with all other fields in $NS$ sector if the following equations are satisfied
\ber
Q^{\vbt_{a}}_{\vl,\vt}-\sum_{i}{\bet_{ai}w_{i}\ov k_{i}+2}\in\mathbb{Z}, \ a=0,...,K-1.
\label{3.NSNSloc}
\enr

Thus, this system of equations singles out the admissible $(\vl,\vt)$ (or $(\vl,\vq)$) of $NS$ primary fields $\Psi^{NS}_{\vl,\vt,\vw}(z,\bar{z})$ which appear in the twisted sector $\vw$ of the orbifold model.

The third step of the construction is simple: we generate $R$ sector fields as
\ber
\Psi^{R}_{\vl,\vt,\vw}(z,\bar{z})=\prod_{i}U^{1\ov 2}_{i}\bar{U}^{1\ov 2}_{i}
\Psi^{NS}_{\vl,\vt,\vw}(z,\bar{z}).
\label{3.RPrimo}
\enr

These fields together with the mutually local $NS$ fields constructed before can be considered as a set of primary fields of the orbifolds in the sense that all the other fields of the orbifold are generated by applying to them the creation operators of the $N=(2,2)$ Virasoro superalgebras of $M_{k_{i}}$ models.

It is clear that $NS$ sector fields of the orbifold are mutually local (by construction), while $R$ sector fields are mutually quasi-local among themselves and with $NS$ sector fields. In other words, the quasi-locality structure of the orbifold is the same as for any $N=(1,1)$ superconformal model.

Using the explicit expressions (\ref{3.Primo}), (\ref{3.RPrimo}) and (\ref{3.NSNSloc}) one can find which of the constructed fields are mutually local. It is not difficult to see that the conditions of total mutual locality are
\be
\begin{aligned}
&Q^{\vbt_{a}}_{\vl,\vt}-\sum_{i}{\bet_{ai}w_{i}\ov k_{i}+2}\in\mathbb{Z}, \quad a=0,...,K-1,
&F+\bar{F}+\sum_{i}w_{i}\in 2\mathbb{Z}.
\label{3.NSRlocal}
\end{aligned}
\ee

These equations single out the set of the  mutually local fields 
for the orbifold model.


\section{Fulfilling the Bootstrap axioms} 
\label{sec:4}

It is easy to see that OPE of the primary fields of the orbifold is closed. Indeed, let us first consider  the OPE of an arbitrary pair of the orbifold fields  $\Psi_{\vl_{1},\vt_{1},\vw_{1}}(z,\bar{z})$ and $\Psi_{\vl_{2},\vt_{2},\vw_{2}}(0)$ from (\ref{3.Primo}), which satisfy (\ref{3.NSNSloc}). 
The charge $Q^{\vbt_{a}}_{\vl,\vt}$ of the resulting field must be equal to the sum of charges $Q^{\vbt_{a}}_{\vl_{1},\vt_{1}}+Q^{\vbt_{a}}_{\vl_{2},\vt_{2}}$ of the original fields  since $U(1)$-charge conservation is performed for each minimal model $M_{k_{i }}$. 
On the other hand, the twist $\vw$ of the resulting field is equal to the sum  the $\vw_{1}+\vw_{2}$ of   twists of the original fields.
Hence, the resulting field will again satisfy the mutual locality equations 
(\ref{3.NSNSloc}). 

These arguments can be directly extended  to include orbifold fields from the $R$ sector (\ref{3.RPrimo}). 
It is also easy to check that the OPE of all (including quasi-local) orbifold fields (\ref{3.Primo}), (\ref{3.RPrimo}) is closed. 
Let us now consider the OPE of a mutually local subsector of the space of all quasilocal fields considered above.
The fields from this sector are selected by the equations 
(\ref{3.NSRlocal}).  
We have just discussed the consistency of OPE with the first equation of this pair.
So let us consider the second equation from (\ref{3.NSRlocal}). This equation can be  rewritten in the form
\ber
\exp{[\im\pi\sum_{i}(J_{0,i}-\bar{J}_{0,i})]}=1,
\enr
because, as one can check, the action of the operator $\exp{[\im\pi\sum_{i}(J_{0,i}-\bar{J}_{0,i})]}$ on any field of the orbifold is given by $\exp{[\im\pi(F+\bar{F}+\sum_{i}w_{i})]}$. 
Since the holomorphic and anti-holomorphic $U(1)$ charges are conserved by OPE, we conclude that the second equation from (\ref{3.NSRlocal}) is consistent with the OPE. 
That is, if both fields entering to  OPE, obey the second equation from (\ref{3.NSRlocal}), then the resulting field also obeys this equation. 
Hence, the OPE for the mutually local subsector of fields is closed.

Using the obtained subset of the strongly  mutually local  fields, 
we  construct the partition function as follows
\ber
Z_{G_{adm}}={1\ov|G_{adm}|}\sum_{\vw\in G_{adm}}\sum_{\vl,\vt}
\prod_{a=0}^{K-1}\dlt(Q^{\vbt_{a}}_{\vl,\vt}-\sum_{i}{\bet_{ai}w_{i}\ov k_{i}+2})
\nmb
(NS_{\vl,\vt+\vw}NS^{*}_{\vl,\vt}+R_{\vl,\vt+\vw}R^{*}_{\vl,\vt}+
\nmb
\exp{[\im\pi(\sum_{i}w_{i})]}(\tld{NS}_{\vl,\vt+\vw}\tld{NS}^{*}_{\vl,\vt}+
\tld{R}_{\vl,\vt+\vw}\tld{R}^{*}_{\vl,\vt})).
\enr

By direct verification, we have seen that this is a modular invariant.  
Thus, the built model satisfies all the requirements of Conformal Bootstrap.
This completes the construction of the $N=2$ SUSY orbifold model.


\section{$(c,c)$ and $(a,c)$ primary fields}
\label{sec:5}

In this section  we use the above  spectral flow construction 
to find $(c,c)$ and $(a,c)$ primary fields.

Before we formulate the algorithms of finding $(c,c)$ and $(a,c)$ fields let's refine the formulas (\ref{2.Prim1}), (\ref{2.Prim2}) to see when our spectral flow construction gives the chiral or anti-chiral primary states for individual Minimal model. We have
\be
\begin{aligned}
&t=0:V_{l,0}(z)=\Phi^{c}_{l}(z)\ \text{is a chiral primary},
\\
&t=l:V_{l,l}(z)=(UG^{-}_{-{1\ov 2}})^{l}\Phi^{c}_{l}(z)\approx\Phi^{a}_{l}(z)\ \text{is an anti-chiral primary},
\\
&t=l+1:V_{l,l+1}(z)=U(UG^{-}_{-{1\ov 2}})^{l}\Phi^{c}_{l}(z)\approx\Phi^{c}_{\tld{l}}(z),\ \tld{l}=k-l,
\ \text{is a chiral primary},
\\
&t=k+1:V_{l,k-l}(z)=(UG^{-}_{-{1\ov 2}})^{k-l}U(UG^{-}_{-{1\ov 2}})^{l}\Phi^{c}_{l}(z)\approx\Phi^{a}_{\tld{l}}(z), 
\,\tld{l}=k-l, \ \text{is an anti-chiral primary}.
\label{5.Mca}
\end{aligned}
\ee

In order to get $(c,c)$ primary field in orbifold model one must have  $(c,c)$ primary field in each  $M_{k_{i}}$ factor of the composite model $M_{\vk}$. Therefore, according to the general formula (\ref{3.Primo}) we can see that in the anti-holomorphic sector the chiral primary state is given by the first line in (\ref{5.Mca}) in each factor $M_{k_{i}}$, while in the holomorphic sector the chiral primary state is given by the first line or by the third line in (\ref{5.Mca}). Taking into account 
(\ref{3.Primo})-(\ref{3.Primo2}), one can deduce the way to build 
the $(c,c)$ fields.
\vskip .3cm

\paragraph{Algorithm for finding $(c,c)$ fields in the twisted $\vw$ sector.}
It consists of the following three steps:\\
{\bf 1.} For each $N=0,1,2,3$ to find all the vectors $\vl=(l_{1},...,l_{5})$ satisfying the equations
\ber
\sum_{i}{l_{i}\ov k_{i}+2}=N, \quad \sum_{i}\bet_{ai}{l_{i}-w_{i}\ov k_{i}+2}\in\mathbb{Z},\quad
a=0,...,K-1.
\label{5.CC1}
\enr
{\bf 2.} Among the found vectors $\vl$, we leave only those for which the set of numbers $w_i$ satisfies the following relations 
\be
\begin{aligned} 
 w_{i}&= 0 \mod\ k_{i}+2, 
\\
& \text{or}
\\
 w_{i}&=l_{i}+1.& 
\label{5.CC2}
\end{aligned}
\ee
{\bf 3.} For the found vectors $\vl$, we define the set of numbers
$\vtl=(\tld{l}_{1},...,\tld{l}_{5})$, built according to the following rules
\be
\begin{aligned}
\tld{l}_{i}=l_{i} \quad &\text{if}\quad \ w_{i}= 0\mod\ k_{i}+2, 
\\
& \text{or}
\\
\tld{l}_{i}=k_{i}-l_{i} \quad &\text{if} \quad w_{i}=l_{i}+1.
\label{5.CC2}
\end{aligned}
\ee
Thus we have got  the pairs $(\vl_{L}=\vtl,\,\vl_{R}=\vl)$ for all $(c,c)$ primary fields of the model.
The holomorphic and anti-holomorphic charges for each of them  are given by
\ber
Q_L=\sum_{i=1}^{5}{\tld{l}_{i}\ov k_{i}+2},\quad {Q_R}=N.
\label{5.QCC}
\enr

Similarly, one can obtain the algorithm of finding $(a,c)$ primaries in the orbifold. In this case one should take $(a,c)$ primary field in each factor $M_{k_{i}}$ of the composite model $M_{\vk}$. Therefore, in the anti-holomorphic sector the chiral primary state is given again by the first line from (\ref{5.Mca}) in each factor $M_{k_{i}}$, while in the holomorphic sector anti-chiral primary state is given by the second line or by the fourth line from (\ref{5.Mca}). Hence, taking into account (\ref{3.Primo})-(\ref{3.Primo2}) we obtain the following algorithm.

\paragraph{Algorithm for finding $(a,c)$ fields in the twisted $\vw$ sector.}
It consists of the following three steps:\\
{\bf 1.} For each $N=0,...,3$ find out all the vectors $\vl=(l_{1},...,l_{5})$ satisfying the equations
\ber
\sum_{i}{l_{i}\ov k_{i}+2}=N, \quad \sum_{i}\bet_{ai}{l_{i}-w_{i}\ov k_{i}+2}\in\mathbb{Z} ,\quad
a=0,...,K-1. 
\label{5.AC1}
\enr
{\bf 2.}  Among the found vectors $\vl$, we keep only those for which the set of numbers $w_i$ satisfies to the following relations
\be
\begin{aligned}
 w_{i}&= l_{i}, 
\\
 &\text{or}
\\
  w_{i}&=k_{i}+1.
\label{5.AC2}
\end{aligned}
\ee
{\bf 3.}  For the found vectors $\vl$, we define the set of numbers $\vtl=(\tld{l}_{1},...,\tld{l}_{5})$, which is built according to the following rules
\be
\begin{aligned}
\tld{l}_{i}=l_{i} \quad&\text{if} \quad w_{i}= l_{i}, 
\\
&\text{or}
\\
\tld{l}_{i}=k_{i}-l_{i} \quad &\text{if} \quad
w_{i}=k_{i}+1.
\label{5.AC2}
\end{aligned}
\ee

Thus we have got the pairs $(\vl_{L}=\vtl,\,\vl_{R}=\vl)$ for $(a,c)$ primary fields of the model.
The holomorphic and anti-holomorphic charges of the $(a,c)$ field are given by
\ber
Q_L=-\sum_{i=1}^{5}{\tld{l}_{i}\ov k_{i}+2},\quad Q_R=N,
\label{5.QAC}
\enr


\section{Examples of $N=2$ orbifolds of Fermat types}
\label{sec:6}

It is known \cite{LVW} that the fields $(c,c)$ and $(a,c)$ are related to the cohomology groups of the corresponding varieties CY if the charges of these fields are correctly identified with the gradings of the cohomology groups. This allows us to identify the fields $(c,c)$ and $(a,c)$ constructed by the algorithms in Section~\ref{sec:5} with members of the Fermat-type CY-orbifold cohomology groups.

Below we give some examples of $N=(2,2)$ orbifold models with $c=9$ and write down explicitly the fields $(c,c)$ and $(a,c)$ with the charges $(1,1)$ and $(-1,1)$ respectively.
We also  show  their relation with $H^{2,1}$ and $H^{1,1}$ cohomology groups of the corresponding CY orbifolds. 

Our examples will be those presented in \cite{GrPl}. Namely, we consider composite model $M_{(3,3,3,3,3)}$ and the orbifold models determined from it  by some admissible groups labeled by their generators $(\vbt_{0},\vbt_{1},...,\vbt_{K-1})$. 
In the corresponding tables which are collected in Appendix~\ref{sec:App} we present the couples of vectors $(\vl_L=(l_{1},...,l_{5}),\vl_R=(\tld{l}_{1},...,\tld{l}_{5}))$, labeling $(c,c)$ or $(a,c)$ fields, according to the twisted sectors where these fields appear. 
To save space, we do not explicitly write out the expressions for these fields, because the reader can easily do it following the rules of algorithms. 

Having our explicit construction of chiral rings and the distribution of these fields among the twisted sectors we can perform more detailed comparison of the constructed fields with the results of calculations within the geometric approach. In \cite{BE} the quotient of  $M_{(3,3,3,3,3)}$ composite model by the admissible group generated by $\vbt_{0}=(1,1,1,1,1), \vbt_{1}=(0,0,0,1,4)$ have been considered in the geometric approach. It was found that only  25 of 49 $(1,1)$-charged $(c,c)$ fields generating polynomial part of $H^{2,1}$ are given by the polynomial monomials, while the rest 24  fields are expressed  in terms of the Laurent polynomials. 

As we can see from the Table~(49,5) in  Appendix~\ref{sec:AppA1}, the spectral flow construction shows the appearance of these 24 $(c,c)$ fields in the twisted sectors. This fact probably explains why these fields could not be found among the polynomials.
 The tables are joined in pairs, so that two mutually mirror manifolds 
from \cite{GrPl} are on the same page.

Using the algorithms of (c,c) and (a,c) rings calculations we calculated dimensions $d_{c,c}^\beta(Q,\bar{Q})$, $d_{a,c}^\beta(Q,\bar{Q})$ of $(Q,\bar{Q})$-charged subspaces of (c,c) and (a,c) rings in all cases considered in 
\cite{GrPl}.  We found complete agreement of these dimensions with the dimensions of the corresponding cohomology groups of mutually mirror CY manifolds. To verify this we use  the relations:  $\dim H^{3-Q,\bar{Q}}=d_{c,c}^\beta(Q,\bar{Q})$ and $ \dim H^{3+Q,\bar{Q}}=d_{a,c}^\beta(Q,\bar{Q})$ \cite{Vafa}.

In the case (49,5) (see Appendix~\ref{sec:AppA1}),  
which is defined by the admissible groups, whose generators are $\vbt_{0}=\{1,1,1,1,1\}$, 
$\vbt_{1}=\{0,0,0,1,4\}$ for the initial orbifold and 
$\vbt_{0}^*=\{1,1,1,1,1\}$, $\vbt_1^*=\{0,1,4,0,0\}$, 
$\vbt_2^*=\{0,3,0,1,1\}$ for its mirror, we got the equalities
$d_{cc}^{\vec{\beta_a}} (1,1)=d_{cc}^{\vec{\beta_a}}(2,2)=49, \  d_{cc}^{\vec{\beta_a}}(2,1)=d_{cc}^{\vec{\beta_a}}(1,2)=5$
 for dimensions of the corresponding  subspaces of the (c,c) ring.
 While for the mirror CY manifold  we got  the same set of dimensions 
$d_{ac}^{\vec{\beta_a^*}} (-1,1)=d_{ac}^{\vec{\beta_a^*}} (-2,2)=49, \  d_{ac}^{\vec{\beta_a^*}} (-2,1)=d_{ac}^{\vec{\beta_a^*}} (-1,2)=5$
 for the  subspaces of the  (a,c)  ring. For more details, see the table below.

The above fact about matching the dimensions of the  cohomology groups for the corresponding CY manifolds  is a manifestation of the mirror symmetry between the two considered $N=2$ SCFT models \cite{LVW, Vafa}.
\\
\\
\begin{table}[ht]
\begin{tabular}{|c:c|c:c|}
\hline
   $d_{cc}^{\vec{\beta_a}} (1,1) =49 $ &  $d_{cc}^{\vec{\beta_a}} (2,1) =5$ &  $d_{cc}^{\vec{\beta^*_a}} (1,1) =5$ &  $d_{cc}^{\vec{\beta^*_a}} (2,1) =49$ \\
      \hdashline
   $d_{cc}^{\vec{\beta_a}} (1,2) =5$ &  $d_{cc}^{\vec{\beta_a}} (2,2) =49$  &  $d_{cc}^{\vec{\beta^*_a}} (1,2) =49$ &  $d_{cc}^{\vec{\beta^*_a}} (2,2) =5$ \\ 
   \hline
   $d_{ac}^{\vec{\beta_a}} (-1,1) =5 $ & $d_{ac}^{\vec{\beta_a}} (-2,1) =49 $ & $d_{ac}^{\vec{\beta^*_a}} (-1,1) =49 $ & $d_{ac}^{\vec{\beta^*_a}} (-2,1) =5 $\\ 
   \hdashline
    $d_{ac}^{\vec{\beta_a}} (-1,2) =49$ &  $d_{ac}^{\vec{\beta_a}} (-2,2) =5$ &  $d_{ac}^{\vec{\beta^*_a}} (-1,2) =5$ &  $d_{ac}^{\vec{\beta^*_a}} (-2,2) =49$ \\
    \hline
\end{tabular}
   \label{table:Symm} 
\end{table}


\section{Conclusion}
\label{sec:7}
In this paper we presented the explicit construction of the fields in a certain class of $N=(2,2)$  superconformal models with $c=9$, which are the orbifolds of the products of $N=(2,2)$ Minimal models. 

We have  shown how to use the products of the exactly solvable
$N=2$ Minimal models, the structure of the Minimal $N=2$ Virasoro superalgebra representations and the spectral flow automorphism to explicitly construct the total set of fields in the orbifold models.

In this way we have obtained a new class of exactly solvable  models of $N=2$ SCFT with a central charge equal to $9$, since these models are defined in terms of the known exactly solvable CFT models, namely, in terms of the parafermionic field theory \cite{ZF} and the theory of free bosonic field.

We explicitly constructed chiral and antichiral rings for several concrete examples of orbifolds and showed that their dimensions agree with the dimensions of the cohomology groups of mutually mirror CY-manifolds. Tables of the results of this calculation are given below in the section ~\ref{sec:App}.
These results agree with those  obtained in  \cite{GrPl} and thus confirm the correctness of our approach.

We note that there are several interesting questions regarding our construction.

First of all, comparing the results of the explicit construction in the table (49,5) with the results in \cite{BE}, we find that the elements of $H^{2,1}$ appearing in twisted sectors cannot be realized as polynomials in the geometric approach.
Therefore, an interesting question arises, what is the geometric interpretation of the $H^{2,1}$ elements that arise in the twisted sectors.

It would be interesting to generalize the considered {\it spectral flow twisting approach}  to include boundary states
and to consider N = (2,2) SCFT of orbifolds with boundaries. This problem has been partially studied already in \cite{P}, but it makes sense to explore such models using the more general constructions developed in this paper.

Finally, it would be interesting to generalize our approach to the cases of  the Landau-Ginzburg $N=2$ orbifold models with the more general Berglund-Hubsch   type superpotentials \cite {BH}, taking into account that some of them connected with Fermat type  $N=2$ orbifold models, as it was shown recently in \cite{BBK}.

\section*{Acknowledgments}

We are gratefull to D. Gepner  for useful discussions.
The work of A. Belavin and S. Parkhomenko is supported by the Russian Science Foundation grant 18-12-00439. 
A. Belavin acknowledges the hospitality of Physics Department at Weizmann Institute of Science where the part of the work was written.

\appendix

\section{Tables}
\label{sec:App}
\subsection{Case (49,5)}
\label{sec:AppA1}

\begin{table}[ht]
\centering
\begin{adjustbox}{width=.9\textwidth}
\small
\begin{tabular}{|| r  r ||r || r  r  r   r  r || r  r  r  r  r || }
\hline
  \multicolumn{13}{||c||}{{\bf c-c, Total number $49$}} \\
   \multicolumn{13}{||c||}{} \\
  \hline
 $n$ & $m$ & cnt & $\tilde{l}_1$ & $\tilde{l}_2$ & $\tilde{l}_3$ & $\tilde{l}_4$ & $\tilde{l}_5$ & $l_1$ & $l_2$ & $l_3$ & $l_4$ &  $l_5$\\ 
  \hline
 0 & 0 & 1 & 3 & 2 & 0 & 0 & 0 & 3 & 2 & 0 & 0 & 0 \\
 0 & 0 & 2 & 2 & 3 & 0 & 0 & 0 & 2 & 3 & 0 & 0 & 0 \\
 0 & 0 & 3 & 3 & 1 & 1 & 0 & 0 & 3 & 1 & 1 & 0 & 0 \\
 0 & 0 & 4 & 2 & 2 & 1 & 0 & 0 & 2 & 2 & 1 & 0 & 0 \\
 0 & 0 & 5 & 1 & 3 & 1 & 0 & 0 & 1 & 3 & 1 & 0 & 0 \\
 0 & 0 & 6 & 3 & 0 & 2 & 0 & 0 & 3 & 0 & 2 & 0 & 0 \\
 0 & 0 & 7 & 2 & 1 & 2 & 0 & 0 & 2 & 1 & 2 & 0 & 0 \\
 0 & 0 & 8 & 1 & 2 & 2 & 0 & 0 & 1 & 2 & 2 & 0 & 0 \\
 0 & 0 & 9 & 0 & 3 & 2 & 0 & 0 & 0 & 3 & 2 & 0 & 0 \\
 0 & 0 & 10 & 2 & 0 & 3 & 0 & 0 & 2 & 0 & 3 & 0 & 0 \\
 0 & 0 & 11 & 1 & 1 & 3 & 0 & 0 & 1 & 1 & 3 & 0 & 0 \\
 0 & 0 & 12 & 0 & 2 & 3 & 0 & 0 & 0 & 2 & 3 & 0 & 0 \\
 0 & 0 & 13 & 3 & 0 & 0 & 1 & 1 & 3 & 0 & 0 & 1 & 1 \\
 0 & 0 & 14 & 2 & 1 & 0 & 1 & 1 & 2 & 1 & 0 & 1 & 1 \\
 0 & 0 & 15 & 1 & 2 & 0 & 1 & 1 & 1 & 2 & 0 & 1 & 1 \\
 0 & 0 & 16 & 0 & 3 & 0 & 1 & 1 & 0 & 3 & 0 & 1 & 1 \\
 0 & 0 & 17 & 2 & 0 & 1 & 1 & 1 & 2 & 0 & 1 & 1 & 1 \\
 0 & 0 & 18 & 1 & 1 & 1 & 1 & 1 & 1 & 1 & 1 & 1 & 1 \\
 0 & 0 & 19 & 0 & 2 & 1 & 1 & 1 & 0 & 2 & 1 & 1 & 1 \\
 0 & 0 & 20 & 1 & 0 & 2 & 1 & 1 & 1 & 0 & 2 & 1 & 1 \\
 0 & 0 & 21 & 0 & 1 & 2 & 1 & 1 & 0 & 1 & 2 & 1 & 1 \\
 0 & 0 & 22 & 0 & 0 & 3 & 1 & 1 & 0 & 0 & 3 & 1 & 1 \\
 0 & 0 & 23 & 1 & 0 & 0 & 2 & 2 & 1 & 0 & 0 & 2 & 2 \\
 0 & 0 & 24 & 0 & 1 & 0 & 2 & 2 & 0 & 1 & 0 & 2 & 2 \\
 0 & 0 & 25 & 0 & 0 & 1 & 2 & 2 & 0 & 0 & 1 & 2 & 2 \\
   \hline
     \hline
 0 & 1 & 1 & 2 & 0 & 0 & 3 & 0 & 2 & 0 & 0 & 0 & 3 \\
 0 & 1 & 2 & 1 & 1 & 0 & 3 & 0 & 1 & 1 & 0 & 0 & 3 \\
 0 & 1 & 3 & 0 & 2 & 0 & 3 & 0 & 0 & 2 & 0 & 0 & 3 \\
 0 & 1 & 4 & 1 & 0 & 1 & 3 & 0 & 1 & 0 & 1 & 0 & 3 \\
 0 & 1 & 5 & 0 & 1 & 1 & 3 & 0 & 0 & 1 & 1 & 0 & 3 \\
 0 & 1 & 6 & 0 & 0 & 2 & 3 & 0 & 0 & 0 & 2 & 0 & 3 \\
   \hline
     \hline
 0 & 2 & 1 & 2 & 0 & 0 & 2 & 1 & 2 & 0 & 0 & 1 & 2 \\
 0 & 2 & 2 & 1 & 1 & 0 & 2 & 1 & 1 & 1 & 0 & 1 & 2 \\
 0 & 2 & 3 & 0 & 2 & 0 & 2 & 1 & 0 & 2 & 0 & 1 & 2 \\
 0 & 2 & 4 & 1 & 0 & 1 & 2 & 1 & 1 & 0 & 1 & 1 & 2 \\
 0 & 2 & 5 & 0 & 1 & 1 & 2 & 1 & 0 & 1 & 1 & 1 & 2 \\
 0 & 2 & 6 & 0 & 0 & 2 & 2 & 1 & 0 & 0 & 2 & 1 & 2 \\
   \hline
     \hline
 0 & 3 & 1 & 2 & 0 & 0 & 1 & 2 & 2 & 0 & 0 & 2 & 1 \\
 0 & 3 & 2 & 1 & 1 & 0 & 1 & 2 & 1 & 1 & 0 & 2 & 1 \\
 0 & 3 & 3 & 0 & 2 & 0 & 1 & 2 & 0 & 2 & 0 & 2 & 1 \\
 0 & 3 & 4 & 1 & 0 & 1 & 1 & 2 & 1 & 0 & 1 & 2 & 1 \\
 0 & 3 & 5 & 0 & 1 & 1 & 1 & 2 & 0 & 1 & 1 & 2 & 1 \\
 0 & 3 & 6 & 0 & 0 & 2 & 1 & 2 & 0 & 0 & 2 & 2 & 1 \\
   \hline
     \hline
 0 & 4 & 1 & 2 & 0 & 0 & 0 & 3 & 2 & 0 & 0 & 3 & 0 \\
 0 & 4 & 2 & 1 & 1 & 0 & 0 & 3 & 1 & 1 & 0 & 3 & 0 \\
 0 & 4 & 3 & 0 & 2 & 0 & 0 & 3 & 0 & 2 & 0 & 3 & 0 \\
 0 & 4 & 4 & 1 & 0 & 1 & 0 & 3 & 1 & 0 & 1 & 3 & 0 \\
 0 & 4 & 5 & 0 & 1 & 1 & 0 & 3 & 0 & 1 & 1 & 3 & 0 \\
 0 & 4 & 6 & 0 & 0 & 2 & 0 & 3 & 0 & 0 & 2 & 3 & 0 \\
\hline
   \multicolumn{13}{c}{} \\
   \multicolumn{13}{c}{} \\
    \hline
  \multicolumn{13}{||c||}{{\bf a-c, Total number $5$}} \\
    \multicolumn{13}{||c||}{} \\
    \hline
 $n$ & $m$ & cnt & $\tilde{l}_1$ & $\tilde{l}_2$ & $\tilde{l}_3$ & $\tilde{l}_4$ & $\tilde{l}_5$ & $l_1$ & $l_2$ & $l_3$ & $l_4$ &  $l_5$\\ 
  \hline
 0 & 2 & 1 & 0 & 0 & 0 & 2 & 3 & 0 & 0 & 0 & 2 & 3 \\
 \hline
 0 & 3 & 1 & 0 & 0 & 0 & 3 & 2 & 0 & 0 & 0 & 3 & 2 \\
 \hline
 1 & 0 & 1 & 1 & 1 & 1 & 1 & 1 & 1 & 1 & 1 & 1 & 1 \\
 \hline
 1 & 1 & 1 & 1 & 1 & 1 & 2 & 0 & 1 & 1 & 1 & 2 & 0 \\
 \hline
 1 & 4 & 1 & 1 & 1 & 1 & 0 & 2 & 1 & 1 & 1 & 0 & 2 \\
\hline
\end{tabular}
\hspace{1cm}
\begin{tabular}{|| r  r  r ||r || r  r  r   r  r || r  r  r  r  r || }
\hline
  \multicolumn{14}{||c||}{{\bf a-c, Total number $49$}} \\
    \multicolumn{14}{||c||}{} \\
    \hline
 $n$ & $m_1$ & $m_2$& cnt & $\tilde{l}_1$ & $\tilde{l}_2$ & $\tilde{l}_3$ & $\tilde{l}_4$ & $\tilde{l}_5$ & $l_1$ & $l_2$ & $l_3$ & $l_4$ &  $l_5$\\ 
  \hline
 0 & 2 & 0 & 1 & 0 & 2 & 3 & 0 & 0 & 0 & 2 & 3 & 0 & 0 \\
   \hline
 0 & 3 & 0 & 1 & 0 & 3 & 2 & 0 & 0 & 0 & 3 & 2 & 0 & 0 \\
   \hline
 0 & 0 & 1 & 1 & 0 & 3 & 0 & 1 & 1 & 0 & 3 & 0 & 1 & 1 \\
   \hline
 0 & 2 & 1 & 1 & 0 & 0 & 3 & 1 & 1 & 0 & 0 & 3 & 1 & 1 \\
   \hline
 0 & 3 & 1 & 1 & 0 & 1 & 2 & 1 & 1 & 0 & 1 & 2 & 1 & 1 \\
   \hline
 0 & 4 & 1 & 1 & 0 & 2 & 1 & 1 & 1 & 0 & 2 & 1 & 1 & 1 \\
   \hline
 0 & 0 & 2 & 1 & 0 & 1 & 0 & 2 & 2 & 0 & 1 & 0 & 2 & 2 \\
   \hline
 0 & 4 & 2 & 1 & 0 & 0 & 1 & 2 & 2 & 0 & 0 & 1 & 2 & 2 \\
   \hline
 1 & 0 & 0 & 1 & 1 & 1 & 1 & 1 & 1 & 1 & 1 & 1 & 1 & 1 \\
   \hline
 1 & 1 & 0 & 1 & 1 & 2 & 0 & 1 & 1 & 1 & 2 & 0 & 1 & 1 \\
   \hline
 1 & 4 & 0 & 1 & 1 & 0 & 2 & 1 & 1 & 1 & 0 & 2 & 1 & 1 \\
   \hline
 1 & 1 & 1 & 1 & 1 & 0 & 0 & 2 & 2 & 1 & 0 & 0 & 2 & 2 \\
   \hline
 1 & 0 & 4 & 1 & 1 & 3 & 1 & 0 & 0 & 1 & 3 & 1 & 0 & 0 \\
   \hline
 1 & 3 & 4 & 1 & 1 & 1 & 3 & 0 & 0 & 1 & 1 & 3 & 0 & 0 \\
   \hline
 1 & 4 & 4 & 1 & 1 & 2 & 2 & 0 & 0 & 1 & 2 & 2 & 0 & 0 \\
   \hline
 2 & 0 & 3 & 1 & 2 & 1 & 2 & 0 & 0 & 2 & 1 & 2 & 0 & 0 \\
   \hline
 2 & 1 & 3 & 1 & 2 & 2 & 1 & 0 & 0 & 2 & 2 & 1 & 0 & 0 \\
   \hline
 2 & 2 & 3 & 1 & 2 & 3 & 0 & 0 & 0 & 2 & 3 & 0 & 0 & 0 \\
   \hline
 2 & 4 & 3 & 1 & 2 & 0 & 3 & 0 & 0 & 2 & 0 & 3 & 0 & 0 \\
   \hline
 2 & 1 & 4 & 1 & 2 & 0 & 1 & 1 & 1 & 2 & 0 & 1 & 1 & 1 \\
   \hline
 2 & 2 & 4 & 1 & 2 & 1 & 0 & 1 & 1 & 2 & 1 & 0 & 1 & 1 \\
   \hline
 3 & 1 & 2 & 1 & 3 & 0 & 2 & 0 & 0 & 3 & 0 & 2 & 0 & 0 \\
   \hline
 3 & 2 & 2 & 1 & 3 & 1 & 1 & 0 & 0 & 3 & 1 & 1 & 0 & 0 \\
   \hline
 3 & 3 & 2 & 1 & 3 & 2 & 0 & 0 & 0 & 3 & 2 & 0 & 0 & 0 \\
   \hline
 3 & 3 & 3 & 1 & 3 & 0 & 0 & 1 & 1 & 3 & 0 & 0 & 1 & 1 \\
  \hline
     \hline
0 & 0 & 4 & 1 & 0 & 2 & 0 & 0 & 3 & 0 & 2 & 0 & 3 & 0 \\
0 & 0 & 4 & 2 & 0 & 2 & 0 & 1 & 2 & 0 & 2 & 0 & 2 & 1 \\
0 & 0 & 4 & 3 & 0 & 2 & 0 & 2 & 1 & 0 & 2 & 0 & 1 & 2 \\
0 & 0 & 4 & 4 & 0 & 2 & 0 & 3 & 0 & 0 & 2 & 0 & 0 & 3 \\
  \hline
     \hline
0 & 3 & 4 & 1 & 0 & 0 & 2 & 0 & 3 & 0 & 0 & 2 & 3 & 0 \\
0 & 3 & 4 & 2 & 0 & 0 & 2 & 1 & 2 & 0 & 0 & 2 & 2 & 1 \\
0 & 3 & 4 & 3 & 0 & 0 & 2 & 2 & 1 & 0 & 0 & 2 & 1 & 2 \\
0 & 3 & 4 & 4 & 0 & 0 & 2 & 3 & 0 & 0 & 0 & 2 & 0 & 3 \\
  \hline
     \hline
0 & 4 & 4 & 1 & 0 & 1 & 1 & 0 & 3 & 0 & 1 & 1 & 3 & 0 \\
0 & 4 & 4 & 2 & 0 & 1 & 1 & 1 & 2 & 0 & 1 & 1 & 2 & 1 \\
0 & 4 & 4 & 3 & 0 & 1 & 1 & 2 & 1 & 0 & 1 & 1 & 1 & 2 \\
0 & 4 & 4 & 4 & 0 & 1 & 1 & 3 & 0 & 0 & 1 & 1 & 0 & 3 \\
 \hline
     \hline

1 & 0 & 3 & 1 & 1 & 0 & 1 & 0 & 3 & 1 & 0 & 1 & 3 & 0 \\
1 & 0 & 3 & 2 & 1 & 0 & 1 & 1 & 2 & 1 & 0 & 1 & 2 & 1 \\
1 & 0 & 3 & 3 & 1 & 0 & 1 & 2 & 1 & 1 & 0 & 1 & 1 & 2 \\
1 & 0 & 3 & 4 & 1 & 0 & 1 & 3 & 0 & 1 & 0 & 1 & 0 & 3 \\
  \hline
     \hline

1 & 1 & 3 & 1 & 1 & 1 & 0 & 0 & 3 & 1 & 1 & 0 & 3 & 0 \\
1 & 1 & 3 & 2 & 1 & 1 & 0 & 1 & 2 & 1 & 1 & 0 & 2 & 1 \\
1 & 1 & 3 & 3 & 1 & 1 & 0 & 2 & 1 & 1 & 1 & 0 & 1 & 2 \\
1 & 1 & 3 & 4 & 1 & 1 & 0 & 3 & 0 & 1 & 1 & 0 & 0 & 3 \\
 \hline
     \hline

2 & 2 & 2 & 1 & 2 & 0 & 0 & 0 & 3 & 2 & 0 & 0 & 3 & 0 \\
2 & 2 & 2 & 2 & 2 & 0 & 0 & 1 & 2 & 2 & 0 & 0 & 2 & 1 \\
2 & 2 & 2 & 3 & 2 & 0 & 0 & 2 & 1 & 2 & 0 & 0 & 1 & 2 \\
2 & 2 & 2 & 4 & 2 & 0 & 0 & 3 & 0 & 2 & 0 & 0 & 0 & 3 \\
 \hline
   \multicolumn{13}{c}{} \\
     \hline

  \multicolumn{14}{||c||}{{\bf c-c, Total number $5$}} \\
    \multicolumn{14}{||c||}{} \\
    \hline
 $n$ & $m_1$ & $m_2$& cnt & $\tilde{l}_1$ & $\tilde{l}_2$ & $\tilde{l}_3$ & $\tilde{l}_4$ & $\tilde{l}_5$ & $l_1$ & $l_2$ & $l_3$ & $l_4$ &  $l_5$\\ 
  \hline
 0 & 0 & 0 & 1 & 1 & 1 & 1 & 2 & 0 & 1 & 1 & 1 & 2 & 0 \\
 0 & 0 & 0 & 2 & 1 & 1 & 1 & 1 & 1 & 1 & 1 & 1 & 1 & 1 \\
 0 & 0 & 0 & 3 & 1 & 1 & 1 & 0 & 2 & 1 & 1 & 1 & 0 & 2 \\
 0 & 0 & 0 & 4 & 0 & 0 & 0 & 3 & 2 & 0 & 0 & 0 & 3 & 2 \\
 0 & 0 & 0 & 5 & 0 & 0 & 0 & 2 & 3 & 0 & 0 & 0 & 2 & 3 \\
\hline
\end{tabular}
\end{adjustbox}
\caption{Left: $\vbt_{1}=\{0,0,0,1,4\}$; Right: its mirror $\vbt_1^*=\{0,1,4,0,0\}$, $\vbt_2^*=\{0,3,0,1,1\}$}
\label{table:Number of fields}
\end{table}

\newpage
\subsection{Case (17,21)}


\begin{table}[ht]
\centering
\begin{adjustbox}{width=.9\textwidth}
\small
\begin{tabular}{|| r  r ||r || r  r  r   r  r || r  r  r  r  r || }
\hline
  \multicolumn{13}{||c||}{{\bf c-c, Total number $17$}} \\
    \multicolumn{13}{||c||}{} \\
  \hline
 $n$ & $m$ & cnt & $\tilde{l}_1$ & $\tilde{l}_2$ & $\tilde{l}_3$ & $\tilde{l}_4$ & $\tilde{l}_5$ & $l_1$ & $l_2$ & $l_3$ & $l_4$ &  $l_5$\\ 
  \hline
 0 & 0 & 1 & 0 & 3 & 2 & 0 & 0 & 0 & 3 & 2 & 0 & 0 \\
 0 & 0 & 2 & 0 & 2 & 3 & 0 & 0 & 0 & 2 & 3 & 0 & 0 \\
 0 & 0 & 3 & 3 & 1 & 0 & 1 & 0 & 3 & 1 & 0 & 1 & 0 \\
 0 & 0 & 4 & 3 & 0 & 1 & 1 & 0 & 3 & 0 & 1 & 1 & 0 \\
 0 & 0 & 5 & 1 & 2 & 0 & 2 & 0 & 1 & 2 & 0 & 2 & 0 \\
 0 & 0 & 6 & 1 & 1 & 1 & 2 & 0 & 1 & 1 & 1 & 2 & 0 \\
 0 & 0 & 7 & 1 & 0 & 2 & 2 & 0 & 1 & 0 & 2 & 2 & 0 \\
 0 & 0 & 8 & 3 & 1 & 0 & 0 & 1 & 3 & 1 & 0 & 0 & 1 \\
 0 & 0 & 9 & 3 & 0 & 1 & 0 & 1 & 3 & 0 & 1 & 0 & 1 \\
 0 & 0 & 10 & 1 & 2 & 0 & 1 & 1 & 1 & 2 & 0 & 1 & 1 \\
 0 & 0 & 11 & 1 & 1 & 1 & 1 & 1 & 1 & 1 & 1 & 1 & 1 \\
 0 & 0 & 12 & 1 & 0 & 2 & 1 & 1 & 1 & 0 & 2 & 1 & 1 \\
 0 & 0 & 13 & 1 & 2 & 0 & 0 & 2 & 1 & 2 & 0 & 0 & 2 \\
 0 & 0 & 14 & 1 & 1 & 1 & 0 & 2 & 1 & 1 & 1 & 0 & 2 \\
 0 & 0 & 15 & 1 & 0 & 2 & 0 & 2 & 1 & 0 & 2 & 0 & 2 \\
 0 & 0 & 16 & 0 & 0 & 0 & 3 & 2 & 0 & 0 & 0 & 3 & 2 \\
 0 & 0 & 17 & 0 & 0 & 0 & 2 & 3 & 0 & 0 & 0 & 2 & 3 \\
\hline
  \multicolumn{13}{c}{} \\
   \multicolumn{13}{c}{} \\
\hline
  \multicolumn{13}{||c||}{{\bf a-c, Total number $21$}} \\
    \multicolumn{13}{||c||}{} \\
    \hline
 $n$ & $m$ & cnt & $\tilde{l}_1$ & $\tilde{l}_2$ & $\tilde{l}_3$ & $\tilde{l}_4$ & $\tilde{l}_5$ & $l_1$ & $l_2$ & $l_3$ & $l_4$ &  $l_5$\\ 
  \hline
 0 & 1 & 1 & 0 & 1 & 1 & 0 & 3 & 0 & 1 & 1 & 3 & 0 \\
 0 & 1 & 2 & 0 & 1 & 1 & 1 & 2 & 0 & 1 & 1 & 2 & 1 \\
 0 & 1 & 3 & 0 & 1 & 1 & 2 & 1 & 0 & 1 & 1 & 1 & 2 \\
 0 & 1 & 4 & 0 & 1 & 1 & 3 & 0 & 0 & 1 & 1 & 0 & 3 \\
   \hline
     \hline
 0 & 4 & 1 & 0 & 0 & 3 & 1 & 1 & 0 & 3 & 0 & 1 & 1 \\
 0 & 4 & 2 & 0 & 1 & 2 & 1 & 1 & 0 & 2 & 1 & 1 & 1 \\
 0 & 4 & 3 & 0 & 2 & 1 & 1 & 1 & 0 & 1 & 2 & 1 & 1 \\
 0 & 4 & 4 & 0 & 3 & 0 & 1 & 1 & 0 & 0 & 3 & 1 & 1 \\
   \hline
     \hline
 1 & 0 & 1 & 1 & 1 & 1 & 1 & 1 & 1 & 1 & 1 & 1 & 1 \\
   \hline
 1 & 1 & 1 & 1 & 2 & 2 & 0 & 0 & 1 & 2 & 2 & 0 & 0 \\
   \hline
 1 & 4 & 1 & 1 & 0 & 0 & 2 & 2 & 1 & 0 & 0 & 2 & 2 \\
   \hline
 2 & 2 & 1 & 2 & 0 & 3 & 0 & 0 & 2 & 3 & 0 & 0 & 0 \\
 2 & 2 & 2 & 2 & 1 & 2 & 0 & 0 & 2 & 2 & 1 & 0 & 0 \\
 2 & 2 & 3 & 2 & 2 & 1 & 0 & 0 & 2 & 1 & 2 & 0 & 0 \\
 2 & 2 & 4 & 2 & 3 & 0 & 0 & 0 & 2 & 0 & 3 & 0 & 0 \\
   \hline
     \hline
 2 & 3 & 1 & 2 & 0 & 0 & 0 & 3 & 2 & 0 & 0 & 3 & 0 \\
 2 & 3 & 2 & 2 & 0 & 0 & 1 & 2 & 2 & 0 & 0 & 2 & 1 \\
 2 & 3 & 3 & 2 & 0 & 0 & 2 & 1 & 2 & 0 & 0 & 1 & 2 \\
 2 & 3 & 4 & 2 & 0 & 0 & 3 & 0 & 2 & 0 & 0 & 0 & 3 \\
   \hline
     \hline
 3 & 2 & 1 & 3 & 0 & 0 & 1 & 1 & 3 & 0 & 0 & 1 & 1 \\
   \hline
 3 & 3 & 1 & 3 & 1 & 1 & 0 & 0 & 3 & 1 & 1 & 0 & 0 \\
\hline
\end{tabular}
\hspace{1cm}
\begin{tabular}{|| r  r r ||r || r  r  r   r  r || r  r  r  r  r || }
\hline
 \multicolumn{14}{||c||}{{\bf a-c, Total number $17$}} \\
    \multicolumn{14}{||c||}{} \\
    \hline
 $n$ & $m_1$ & $m_2$ & cnt & $\tilde{l}_1$ & $\tilde{l}_2$ & $\tilde{l}_3$ & $\tilde{l}_4$ & $\tilde{l}_5$ & $l_1$ & $l_2$ & $l_3$ & $l_4$ &  $l_5$\\ 
  \hline
 0 & 2 & 0 & 1 & 0 & 2 & 3 & 0 & 0 & 0 & 2 & 3 & 0 & 0 \\
  \hline
 0 & 3 & 0 & 1 & 0 & 3 & 2 & 0 & 0 & 0 & 3 & 2 & 0 & 0 \\
  \hline
 0 & 0 & 2 & 1 & 0 & 0 & 0 & 2 & 3 & 0 & 0 & 0 & 2 & 3 \\
  \hline
 0 & 0 & 3 & 1 & 0 & 0 & 0 & 3 & 2 & 0 & 0 & 0 & 3 & 2 \\
  \hline
 1 & 0 & 0 & 1 & 1 & 1 & 1 & 1 & 1 & 1 & 1 & 1 & 1 & 1 \\
  \hline
 1 & 1 & 0 & 1 & 1 & 2 & 0 & 1 & 1 & 1 & 2 & 0 & 1 & 1 \\
  \hline
 1 & 4 & 0 & 1 & 1 & 0 & 2 & 1 & 1 & 1 & 0 & 2 & 1 & 1 \\
  \hline
 1 & 0 & 1 & 1 & 1 & 1 & 1 & 2 & 0 & 1 & 1 & 1 & 2 & 0 \\
  \hline
 1 & 1 & 1 & 1 & 1 & 2 & 0 & 2 & 0 & 1 & 2 & 0 & 2 & 0 \\
  \hline
 1 & 4 & 1 & 1 & 1 & 0 & 2 & 2 & 0 & 1 & 0 & 2 & 2 & 0 \\
  \hline
 1 & 0 & 4 & 1 & 1 & 1 & 1 & 0 & 2 & 1 & 1 & 1 & 0 & 2 \\
  \hline
 1 & 1 & 4 & 1 & 1 & 2 & 0 & 0 & 2 & 1 & 2 & 0 & 0 & 2 \\
  \hline
 1 & 4 & 4 & 1 & 1 & 0 & 2 & 0 & 2 & 1 & 0 & 2 & 0 & 2 \\
  \hline
 3 & 2 & 2 & 1 & 3 & 0 & 1 & 0 & 1 & 3 & 0 & 1 & 0 & 1 \\
  \hline
 3 & 3 & 2 & 1 & 3 & 1 & 0 & 0 & 1 & 3 & 1 & 0 & 0 & 1 \\
  \hline
 3 & 2 & 3 & 1 & 3 & 0 & 1 & 1 & 0 & 3 & 0 & 1 & 1 & 0 \\
  \hline
 3 & 3 & 3 & 1 & 3 & 1 & 0 & 1 & 0 & 3 & 1 & 0 & 1 & 0 \\
  \hline
  \multicolumn{13}{c}{} \\
   \multicolumn{13}{c}{} \\
\hline
  \multicolumn{14}{||c||}{{\bf c-c, Total number $21$}} \\
    \multicolumn{14}{||c||}{} \\
    \hline
 $n$ & $m_1$ & $m_2$ & cnt & $\tilde{l}_1$ & $\tilde{l}_2$ & $\tilde{l}_3$ & $\tilde{l}_4$ & $\tilde{l}_5$ & $l_1$ & $l_2$ & $l_3$ & $l_4$ &  $l_5$\\ 
  \hline
 0 & 0 & 0 & 1 & 3 & 1 & 1 & 0 & 0 & 3 & 1 & 1 & 0 & 0 \\
 0 & 0 & 0 & 2 & 1 & 2 & 2 & 0 & 0 & 1 & 2 & 2 & 0 & 0 \\
 0 & 0 & 0 & 3 & 3 & 0 & 0 & 1 & 1 & 3 & 0 & 0 & 1 & 1 \\
 0 & 0 & 0 & 4 & 1 & 1 & 1 & 1 & 1 & 1 & 1 & 1 & 1 & 1 \\
 0 & 0 & 0 & 5 & 1 & 0 & 0 & 2 & 2 & 1 & 0 & 0 & 2 & 2 \\
   \hline
     \hline
 0 & 1 & 0 & 1 & 2 & 3 & 0 & 0 & 0 & 2 & 0 & 3 & 0 & 0 \\
 0 & 1 & 0 & 2 & 0 & 3 & 0 & 1 & 1 & 0 & 0 & 3 & 1 & 1 \\
   \hline
     \hline
 0 & 2 & 0 & 1 & 2 & 2 & 1 & 0 & 0 & 2 & 1 & 2 & 0 & 0 \\
 0 & 2 & 0 & 2 & 0 & 2 & 1 & 1 & 1 & 0 & 1 & 2 & 1 & 1 \\
   \hline
     \hline
 0 & 3 & 0 & 1 & 2 & 1 & 2 & 0 & 0 & 2 & 2 & 1 & 0 & 0 \\
 0 & 3 & 0 & 2 & 0 & 1 & 2 & 1 & 1 & 0 & 2 & 1 & 1 & 1 \\  
   \hline
     \hline
 0 & 4 & 0 & 1 & 2 & 0 & 3 & 0 & 0 & 2 & 3 & 0 & 0 & 0 \\
 0 & 4 & 0 & 2 & 0 & 0 & 3 & 1 & 1 & 0 & 3 & 0 & 1 & 1 \\
   \hline
     \hline
 0 & 0 & 1 & 1 & 2 & 0 & 0 & 3 & 0 & 2 & 0 & 0 & 0 & 3 \\
 0 & 0 & 1 & 2 & 0 & 1 & 1 & 3 & 0 & 0 & 1 & 1 & 0 & 3 \\
   \hline
     \hline
 0 & 0 & 2 & 1 & 2 & 0 & 0 & 2 & 1 & 2 & 0 & 0 & 1 & 2 \\
 0 & 0 & 2 & 2 & 0 & 1 & 1 & 2 & 1 & 0 & 1 & 1 & 1 & 2 \\
   \hline
     \hline
 0 & 0 & 3 & 1 & 2 & 0 & 0 & 1 & 2 & 2 & 0 & 0 & 2 & 1 \\
 0 & 0 & 3 & 2 & 0 & 1 & 1 & 1 & 2 & 0 & 1 & 1 & 2 & 1 \\
   \hline
     \hline
 0 & 0 & 4 & 1 & 2 & 0 & 0 & 0 & 3 & 2 & 0 & 0 & 3 & 0 \\
 0 & 0 & 4 & 2 & 0 & 1 & 1 & 0 & 3 & 0 & 1 & 1 & 3 & 0 \\
\hline
\end{tabular}
\end{adjustbox}
\caption{Left: $\vbt_{1}=\{0,1,1,4,4\}$; Right: mirror $\beta_1^*=\{0,1,4,0,0\}$, $\beta_2^*=\{0,0,0,1,4\}$}
\label{table:Number of fields}
\end{table}

\newpage
\subsection{Case (21,1)}

\begin{table}[ht]
\centering
\begin{adjustbox}{width=.9\textwidth}
\small
\begin{tabular}{|| r  r  ||r || r  r  r   r  r || r  r  r  r  r || }
\hline
  \multicolumn{13}{||c||}{{\bf c-c, Total number $21$}} \\
    \multicolumn{13}{||c||}{} \\
  \hline
 $n$ & $m$ & cnt & $\tilde{l}_1$ & $\tilde{l}_2$ & $\tilde{l}_3$ & $\tilde{l}_4$ & $\tilde{l}_5$ & $l_1$ & $l_2$ & $l_3$ & $l_4$ &  $l_5$\\ 
  \hline
 0 & 0 & 1 & 1 & 3 & 1 & 0 & 0 & 1 & 3 & 1 & 0 & 0 \\
 0 & 0 & 2 & 2 & 1 & 2 & 0 & 0 & 2 & 1 & 2 & 0 & 0 \\
 0 & 0 & 3 & 2 & 2 & 0 & 1 & 0 & 2 & 2 & 0 & 1 & 0 \\
 0 & 0 & 4 & 3 & 0 & 1 & 1 & 0 & 3 & 0 & 1 & 1 & 0 \\
 0 & 0 & 5 & 0 & 1 & 3 & 1 & 0 & 0 & 1 & 3 & 1 & 0 \\
 0 & 0 & 6 & 0 & 2 & 1 & 2 & 0 & 0 & 2 & 1 & 2 & 0 \\
 0 & 0 & 7 & 1 & 0 & 2 & 2 & 0 & 1 & 0 & 2 & 2 & 0 \\
 0 & 0 & 8 & 1 & 1 & 0 & 3 & 0 & 1 & 1 & 0 & 3 & 0 \\
 0 & 0 & 9 & 3 & 1 & 0 & 0 & 1 & 3 & 1 & 0 & 0 & 1 \\
 0 & 0 & 10 & 0 & 2 & 2 & 0 & 1 & 0 & 2 & 2 & 0 & 1 \\
 0 & 0 & 11 & 1 & 0 & 3 & 0 & 1 & 1 & 0 & 3 & 0 & 1 \\
 0 & 0 & 12 & 0 & 3 & 0 & 1 & 1 & 0 & 3 & 0 & 1 & 1 \\
 0 & 0 & 13 & 1 & 1 & 1 & 1 & 1 & 1 & 1 & 1 & 1 & 1 \\
 0 & 0 & 14 & 2 & 0 & 0 & 2 & 1 & 2 & 0 & 0 & 2 & 1 \\
 0 & 0 & 15 & 0 & 0 & 1 & 3 & 1 & 0 & 0 & 1 & 3 & 1 \\
 0 & 0 & 16 & 1 & 2 & 0 & 0 & 2 & 1 & 2 & 0 & 0 & 2 \\
 0 & 0 & 17 & 2 & 0 & 1 & 0 & 2 & 2 & 0 & 1 & 0 & 2 \\
 0 & 0 & 18 & 0 & 0 & 2 & 1 & 2 & 0 & 0 & 2 & 1 & 2 \\
 0 & 0 & 19 & 0 & 1 & 0 & 2 & 2 & 0 & 1 & 0 & 2 & 2 \\
 0 & 0 & 20 & 0 & 1 & 1 & 0 & 3 & 0 & 1 & 1 & 0 & 3 \\
 0 & 0 & 21 & 1 & 0 & 0 & 1 & 3 & 1 & 0 & 0 & 1 & 3 \\
\hline
  \multicolumn{13}{c}{} \\
   \multicolumn{13}{c}{} \\
\hline
  \multicolumn{13}{||c||}{{\bf a-c, Total number $1$}} \\
    \multicolumn{13}{||c||}{} \\
    \hline
 $n$ & $m$ & cnt & $\tilde{l}_1$ & $\tilde{l}_2$ & $\tilde{l}_3$ & $\tilde{l}_4$ & $\tilde{l}_5$ & $l_1$ & $l_2$ & $l_3$ & $l_4$ &  $l_5$\\ 
  \hline
 1 & 0 & 1 & 1 & 1 & 1 & 1 & 1 & 1 & 1 & 1 & 1 & 1 \\
\hline
\end{tabular}
\hspace{1cm}
\begin{tabular}{|| r  r r ||r || r  r  r   r  r || r  r  r  r  r || }
 \hline
  \multicolumn{14}{||c||}{{\bf a-c, Total number $21$}} \\
    \multicolumn{14}{||c||}{} \\
    \hline
 $n$ & $m_1$ & $m_2$  & cnt & $\tilde{l}_1$ & $\tilde{l}_2$ & $\tilde{l}_3$ & $\tilde{l}_4$ & $\tilde{l}_5$ & $l_1$ & $l_2$ & $l_3$ & $l_4$ &  $l_5$\\ 
\hline
 0 & 1 & 0 & 1 & 0 & 1 & 3 & 1 & 0 & 0 & 1 & 3 & 1 & 0 \\
 \hline
 0 & 2 & 0 & 1 & 0 & 2 & 1 & 2 & 0 & 0 & 2 & 1 & 2 & 0 \\
 \hline
 0 & 0 & 1 & 1 & 0 & 1 & 1 & 0 & 3 & 0 & 1 & 1 & 0 & 3 \\
 \hline
 0 & 0 & 2 & 1 & 0 & 2 & 2 & 0 & 1 & 0 & 2 & 2 & 0 & 1 \\
 \hline
 0 & 1 & 2 & 1 & 0 & 3 & 0 & 1 & 1 & 0 & 3 & 0 & 1 & 1 \\
 \hline
 0 & 3 & 2 & 1 & 0 & 0 & 1 & 3 & 1 & 0 & 0 & 1 & 3 & 1 \\
 \hline
 0 & 1 & 4 & 1 & 0 & 0 & 2 & 1 & 2 & 0 & 0 & 2 & 1 & 2 \\
 \hline
 0 & 2 & 4 & 1 & 0 & 1 & 0 & 2 & 2 & 0 & 1 & 0 & 2 & 2 \\
 \hline
 1 & 0 & 0 & 1 & 1 & 1 & 1 & 1 & 1 & 1 & 1 & 1 & 1 & 1 \\
 \hline
 1 & 4 & 0 & 1 & 1 & 0 & 3 & 0 & 1 & 1 & 0 & 3 & 0 & 1 \\
 \hline
 1 & 4 & 2 & 1 & 1 & 2 & 0 & 0 & 2 & 1 & 2 & 0 & 0 & 2 \\
 \hline
 1 & 1 & 3 & 1 & 1 & 0 & 2 & 2 & 0 & 1 & 0 & 2 & 2 & 0 \\
 \hline
 1 & 2 & 3 & 1 & 1 & 1 & 0 & 3 & 0 & 1 & 1 & 0 & 3 & 0 \\
 \hline
 1 & 4 & 3 & 1 & 1 & 3 & 1 & 0 & 0 & 1 & 3 & 1 & 0 & 0 \\
 \hline
 1 & 0 & 4 & 1 & 1 & 0 & 0 & 1 & 3 & 1 & 0 & 0 & 1 & 3 \\
 \hline
 2 & 3 & 0 & 1 & 2 & 0 & 1 & 0 & 2 & 2 & 0 & 1 & 0 & 2 \\
 \hline
 2 & 3 & 1 & 1 & 2 & 1 & 2 & 0 & 0 & 2 & 1 & 2 & 0 & 0 \\
 \hline
 2 & 4 & 1 & 1 & 2 & 2 & 0 & 1 & 0 & 2 & 2 & 0 & 1 & 0 \\
 \hline
 2 & 0 & 3 & 1 & 2 & 0 & 0 & 2 & 1 & 2 & 0 & 0 & 2 & 1 \\
 \hline
 3 & 2 & 1 & 1 & 3 & 1 & 0 & 0 & 1 & 3 & 1 & 0 & 0 & 1 \\
 \hline
 3 & 3 & 4 & 1 & 3 & 0 & 1 & 1 & 0 & 3 & 0 & 1 & 1 & 0 \\
 \hline
   \multicolumn{13}{c}{} \\
   \multicolumn{13}{c}{} \\
\hline
\multicolumn{14}{||c||}{{\bf c-c, Total number $1$}} \\
    \multicolumn{14}{||c||}{} \\
    \hline
 $n$ & $m_1$ & $m_2$  & cnt & $\tilde{l}_1$ & $\tilde{l}_2$ & $\tilde{l}_3$ & $\tilde{l}_4$ & $\tilde{l}_5$ & $l_1$ & $l_2$ & $l_3$ & $l_4$ &  $l_5$\\ 
  \hline
 0 & 0 & 0 & 1 & 1 & 1 & 1 & 1 & 1 & 1 & 1 & 1 & 1 & 1 \\
 \hline
\end{tabular}
\end{adjustbox}
\caption{Left: $\vbt_{1}=\{0,1,2,3,4\}$; Right: mirror $\vbt_1^*=\{0,1,3,1,0\}$, $\vbt_2^*=\{0,1,1,0,3\}$}
\label{table:Number of fields}
\end{table}

\end{document}